\UseRawInputEncoding
\documentclass{aa}  
\usepackage{ulem}
\usepackage{graphicx}
\usepackage{txfonts}
\usepackage[pdftex,breaklinks,colorlinks,citecolor=blue,linkcolor=blue]{hyperref}
\usepackage[colorinlistoftodos]{todonotes}

\usepackage{subfig}
\usepackage{siunitx}
\usepackage{adjustbox}

\usepackage{color, colortbl}

\newcommand{\aminvsg}{$a_{\mathrm{min,\,a-C}}$}
\newcommand{\abvsg}{M$_{\mathrm{a-C}}$/M$_{\mathrm{H}}$}
\newcommand{\lpdr}{$l_{\rm{PDR}}$}

\newcommand{\mum}{$\mu$m}
\newcommand{\cm}{a-C}

\newcommand{\CM}{a-C:H/a-C}
\newcommand{\py}{a-Sil/a-C}

\newcommand{\G}{$G_{0}$}
\newcommand{\nh}{$n_{\mathrm{H}}$}

\newcommand{\MIPSun}{$\mathrm{MIPS}_{24}$}
\newcommand{\MIPSdeux}{$\mathrm{PACS}_{70}$}

\newcommand{\IRACun}{$\mathrm{IRAC}_{3.6}$}
\newcommand{\IRACdeux}{$\mathrm{IRAC}_{4.5}$}
\newcommand{\IRACtrois}{$\mathrm{IRAC}_{5.8}$}
\newcommand{\IRACquatre}{$\mathrm{IRAC}_{8.0}$}

\newcommand{\herschel}{\textit{Herschel}}
\newcommand{\spitzer}{\textit{Spitzer}}

\newcommand{\chiMIR}{$\chi_{\mathrm{MIR}}^{2}$}
\newcommand{\chiTOT}{$\chi_{\mathrm{tot}}^{2}$}
   
\newcommand{\OI}{O\,{\sc i}}

\newcommand{\CII}{C\,{\sc ii}}

\newcommand{\HII}{H\,{\sc ii}}

\begin{document}

   \title{Nano-grain depletion in photon-dominated regions}
   
   \subtitle{}

   \author{T. Schirmer.
          \inst{1,2} 
          \and
          N. Ysard\inst{2}
          \and
          E. Habart\inst{2}
          \and
          A. P. Jones\inst{2}
          \and
          A. Abergel\inst{2}
          \and
          L. Verstraete\inst{2}
          }

   \institute{Department of Space, Earth and Environment, Chalmers University of Technology, Onsala Space Observatory, 439 92 Onsala, Sweden \\
   \email{thisch@chalmers.se}
   \and 
   Université Paris-Saclay, CNRS,  Institut d'astrophysique spatiale, 91405, Orsay, France 
          }
   \date{Received 25 March 2022; accepted 27 August 2022}

 
  \abstract
   {Carbonaceous nano-grains play a fundamental role in the physico-chemistry of the interstellar medium (ISM) and especially of photon-dominated regions (PDRs). Their properties vary with the local physical conditions and affect the local chemistry and dynamics.}
   {We aim to highlight the evolution of carbonaceous nano-grains in three different PDRs and propose a scenario of dust evolution as a response to the physical conditions.}
   {We used \spitzer/IRAC (3.6, 4.5, 5.8, and 8 \mum) and \spitzer/MIPS (24 \mum) together with \herschel/PACS (70 \mum) to map dust emission in IC63 and the Orion Bar. To assess the dust properties, we modelled the dust emission in these regions using the radiative transfer code SOC together with the THEMIS dust model.}
   {Regardless of the PDR, we find that nano-grains are depleted and that their minimum size is larger than in the diffuse ISM (DISM), which suggests that the mechanisms that lead nano-grains to be photo-destroyed are very efficient below a given critical size limit. The evolution of the nano-grain dust-to-gas mass ratio with both \G~and the effective temperature of the illuminating star indicates a competition between the nano-grain formation through the fragmentation of larger grains and nano-grain photo-destruction. We modelled dust collisions driven by radiative pressure with a classical 1D approach to show that this is a viable scenario for explaining nano-grain formation through fragmentation and, thus, the variations observed in nano-grain dust-to-gas mass ratios from one PDR to another.}
   {We find a broad variation in the nano-grain dust properties from one PDR to another, along with a general trend of nano-grain depletion in these regions. We propose a viable scenario of nano-grain formation through fragmentation of large grains due to radiative pressure-induced collisions.}


   \maketitle

\section{Introduction}\label{sect:introduction}

Interstellar dust is ubiquitous in the interstellar medium (ISM) and it is
involved into the physical, chemical, and dynamical evolution of numerous environments through different processes such as the gas heating through the photoelectric effect \citep{bakes_photoelectric_1994, weingartner_photoelectric_2001} and the H$_2$ formation on dust surfaces \citep[e.g][]{le_bourlot_surface_2012, bron_stochastic_2014, jones_h_2015}. The efficiency
of these processes depends crucially on the dust properties (size, composition, and shape). It is 
therefore crucial to constrain those properties in order to understand better 
the different environments where dust exists. However, the broad disparity in 
the physical conditions (density and irradiation) triggers an evolution of these dust properties through grain growth (i.e. accretion and coagulation), grain destruction (i.e. photo-destruction and collisions), and processing (i.e. aromatisation, dehydrogenation), which are not yet fully understood.

The radiative feedback of freshly formed stars irradiating their nearby dense environments leads to the creation of the well-known photon-dominated regions (PDRs). In these regions, physical conditions vary significantly on small spatial scales which is why PDRs are a unique place to study how dust evolves as a response to the physical conditions. The mid-IR spectra of PDRs present a wealth of emission band features due to the smallest grains overlying the continuum of hot dust emission which has been extensively analysed using the Infrared Space Observatory (ISO) and Spitzer data. Strong variations in the spectra have been found across and between PDRs \citep[e.g.][]{peeters_rich_2002, peeters_polycyclic_2004,rapacioli_formation_2006, abergel_evolution_2002, berne_analysis_2007}. With Herschel data in the far-IR (FIR), it has also become possible to study the emission of large grains in thermal equilibrium \citep[e.g.][]{abergel_evolution_2010,arab_evolution_2012}. Using the dust model THEMIS \citep{jones_evolution_2013, jones_global_2017} together with the 3D radiative transfer code SOC \citep{juvela_soc_2019}, the study from \cite{schirmer_dust_2020} (hereafter called Paper I) reported that the nano-grain dust-to-gas mass ratio in the irradiated outer part of the Horsehead is 6-10 times lower than in the diffuse ISM and that the minimum size of these grains is 2–2.25 times larger than in the diffuse ISM.

The gas physics and chemistry in PDRs are strongly affected by variations in the dust properties \citep[][]{schirmer_influence_2021}. The aim of this study is thus to constrain nano-grain dust properties in the Orion Bar and in IC63, two PDRs that present very contrasted physical conditions, using \spitzer~and \herschel~observations. Evidence of dust evolution in those two PDRs has already been demonstrated \citep[e.g.][]{arab_evolution_2012, van_de_putte_evidence_2019} and our goal in this work is to understand the different mechanisms that lead to the formation and destruction of nano-grains in these regions.  

The paper is organised as follows. In Sect.\,\ref{sect:observations}, we describe the previous studies of the Orion Bar and IC63. We also present the \spitzer~and \herschel~observations. In Sect.\,\ref{sect:models_and_methods}, we detail the THEMIS dust model as well as the radiative transfer code used to compute dust emission. In Sect.\,\ref{sect:density_IC63_orion}, we present the density profiles and our methodology to constrain those profiles. In Sect.\,\ref{sect:dust_prop}, we compare the dust modelled emission to the observations and the constraints on the dust properties that result from this modelling. In Sect.\,\ref{sect:discussion}, we discuss our results and present a scenario of dust evolution within these regions, based on processing timescales. Our conclusions are given in Sect.\,\ref{sect:conclusion}.

\section{Selected PDRs}\label{sect:observations}

\subsection{Orion Bar}\label{sect:sect:orion}

The Orion Bar (see Fig.\,\ref{fig:IC63_orion_obs_all0}, bottom panel) is a bright filament of the Orion molecular cloud, a site located at 414 pc \citep{menten_distance_2007} that is undergoing massive star-formation. The bar is illuminated by the \mbox{O7-type} star \mbox{$\theta^1$ Ori C}, the most massive member of the Trapezium young stellar cluster,
 at the heart
of the Orion Nebula \citep[about 2$'$ north east of the Bar, e.g.,][]{odell_orion_2001}. The intense ionizing radiation and strong winds from \mbox{$\theta$ Ori C} power and shape the nebula \citep[][]{pabst_disruption_2019,gudel_million-degree_2008}, a blister \HII~region  that is eating its way into the background parental cloud Orion Molecular Cloud (OMC). The incident UV radiation field is \mbox{(1-3)\,$\times $10$^4$} times the mean
interstellar field \citep[e.g.,][]{marconi_near_1998}. The first PDR layers are predominantly neutral and atomic: \mbox{[H]\,$>$\,[H$_2$]\,$\gg$\,[H$^+$]}. They display a wealth of near-infrared (NIR) atomic lines from low ionization potential elements 
\citep[forbidden lines, recombination lines, etc.; see][]{walmsley_structure_2000}.  
Gas is mainly heated by photoelectrons ejected from small grains and it is mostly cooled by the FIR [\CII]\,158\,$\mu$m and [\OI]\,63\,$\mu$m fine-structure lines \citep[e.g.][]{tielens_anatomy_1993,herrmann_orion_1997,bernard-salas_spatial_2012,ossenkopf_herschel_2013}.
The observed narrow (\mbox{$\Delta v$\,=\,2--3\,km\,s$^{-1}$}) carbon and sulfur radio recombination lines also arise from these layers (not from
the \HII~region) and provide a measure of the electron density in the PDR
\citep[$n_{\rm e}$\,$\simeq$\,10--100\,cm$^{-3}$; e.g.][]{wyrowski_carbon_1997,cuadrado_direct_2019,goicoechea_bottlenecks_2021}.
The atomic PDR zone also hosts
the peak of the mid-infrared (MIR) polycyclic aromatic hydrocarbons (PAH) emission \citep[e.g.][]{bregman_infrared_1989,sellgren_33_1990,tielens_anatomy_1993,giard_pah_1994,knight_tracing_2021}, which has led to numerous studies on dust in the Orion Bar. Indeed, based on ISO spectroscopy with the SWS spectrograph of the Orion Bar, \cite{peeters_rich_2002} showed that the variations in the spectral features in the $6-9$ \mum~spectral range (emission bands at $6.0$, $6.2$, $6.6$, $7.0$, $7.7$, $8.3$, and $8.6$ \mum) are linked to variations 
in the local physical conditions but also with both formation and evolution processes associated with PAHs. From imaging and spectroscopic observations with ISOCAM ($5-16$ \mum), \cite{abergel_evolution_2002} showed that at the illuminated edge of the Orion Bar, there is a systematic decrease in the intensity of the aromatic features (especially at $7.7$ \mum) relative to the continuum inside the shielded molecular regions compared to the photo-dissociated and photo-ionised regions. Later, \cite{arab_evolution_2012} showed that a decrease in the PAH abundance together with an increase in the larger grain emissivity, fits the Spitzer and Herschel observations of the Orion Bar well. This can be explained via the scenario of PAHs being photodestroyed or coagulating onto the surface of larger grains, thus increasing their emissivity. Mid-IR photometry with SOFIA of the Orion Bar and the Orion HII region \citep{salgado_orion_2016} requires coagulated grains to explain the decrease by a factor  of $5$ to $10$ of the UV and infrared dust opacities from the diffuse ISM to these PDRs. Using SOFIA, \spitzer, UKIRT, and ISO observations together with the dust destruction model SHIVA \citep{murga_shiva_2019}, \cite{murga_orion_2021} showed that small PAHs ($N_{\mathrm{C}}<60$) are most likely destroyed.   

From now on, we follow \cite{arab_evolution_2012} and adopt \G~= $2.6\times 10^{4}$ where:

\begin{equation}
    G_0=\frac{1}{1.6\times 10^{-3}\,(\mathrm{erg\,s^{-1}\,cm^{-2}})}\int_{6\,\mathrm{eV}}^{13.6\,\mathrm{eV}}I_{\nu}\,d\nu. 
\end{equation}

\subsection{IC63}\label{sect:sect:IC63}

Since it was first discovered with the telescope of the Mount Wilson and Palomar Observatories then mentioned in \cite{sharpless_catalogue_1953}, IC63 (see Fig.\,\ref{fig:IC63_orion_obs_all0}, top panel) has been the subject of various studies \citep[e.g.][]{fleming_spitzer_2010, andersson_evidence_2013, andrews_whipping_2018, dennis_physics_2020,lai_are_2020,soam_collisional_2021,soam_interstellar_2021}. At the very beginning, \cite{witt_uv_1989} found evidence of Extended Red Emission (ERE) in IC63, based on low-resolution UV (115-195 nm) spectra obtained with the International Ultraviolet Explorer (IUE) together with spectra obtained at the McGraw-Hill Observatory, which covers a spectral range from 500 to 900 nm. They estimated that the UV field hitting the surface of IC63 is about $\sim 2.3 \times 10^{5}$ photons\,cm\,s\,Hz$^{-1}$ at 100 nm, hence\footnote{The incident UV radiation field is there expressed in units $\chi$ of the \cite{draine_photoelectric_1978} average interstellar radiation field.} $\chi \sim 680$ (i.e. \G\,$\sim$ 1160). Using this $\chi$ value together with the IUE observations, \cite{sternberg_ultraviolet_1989} found an average density of $n\sim 4\times 10^{4}$ H\,cm$^{-3}$ in this nebula. Based on millimeter and submillimeter observations\footnote{Using the CSO (Caltech Submillimeter Observatory) telescope at Mauna Kea, the NRAO (National Radio Astronomical Observatory) 12 meter telescope at Kitt Peak, the IRAM (Institut de Radio Astronomie Millimétrique) 30 meter telescope at Pico Valeta and the JCMT (James Clerk Maxwell Telescope) 15 meter telescope at Mauna Kea.}, \cite{jansen_physical_1994} constrained the UV field as well as the average density in IC63 using HCO$^+$, CS, HCN, and H$_2$CO lines. They found that $\chi$ is about 650 (i.e. \G\,$\sim$ 1110) and $n(\mathrm{H_{2}})\sim (5\pm 2) \, \times 10^{4}$ cm$^{-3}$ ($n_{\mathrm{H}}\sim 2\,n(\mathrm{H_2})\sim 1\pm0.4 \, \times 10^{5}$ H\,cm$^{-3}$). These results where later reinforced on the basis of chemical calculations for various physical conditions with \nh~varying from $6\times 10^{4}$ and $14\times 10^{4}$ H\,cm$^{-3}$, \cite{jansen_physical_1995} found a best fit corresponding to \nh~= $1\times 10^{5}$ H\,cm$^{-3}$. This value was confirmed in \cite{jansen_physical_1996} and they showed that the carbon abundance in gas-phase is $X_{\mathrm{C}}=(13^{+6}_{-4})$ \% although it is about 30-60 \% in diffuse clouds such as $\zeta$ Oph \citep{cardelli_abundance_1993}. Using H$_2$ pure-rotational lines with the Short Wavelength Spectrometer (SWS) onboard the Infrared Space Observatory (ISO) together with major fine-structure cooling lines of OI at 63 \mum~and 145 \mum~as well as CII at 157.7 \mum,  \cite{thi_weak_1999} found that the incident radiation field has a \G~of about $10^{3}$, which falls in the range of the \G~values found in previous studies. Observations at IRAM 30-m of C$_2$H, c-C$_3$H$_2$, C$_4$H, l-C$_3$H, c-C$_3$H, and HN$^{13}$C with relatively high abundance might support an in situ formation of carbon chains and rings assisted by a release of acetylene from very small carbon particles \citep{fosse_aibs_2000, teyssier_carbon_2004} or the release of C$_n$H$_m$ species from the photo-processing of a-C(:H) nano-particles \citep{jones_h_2015}. Based on extinction mapping, evidence for dust evolution has been shown in IC63 \citep{van_de_putte_evidence_2019}.
As \G~estimates vary between \G~$\sim 1000$ \citep{thi_weak_1999} and \G~$\sim 1200$ \citep{witt_uv_1989, jansen_physical_1994}, we chose \G~= 1100.

\subsection{Ancillary data used in this study}\label{sect:sect:spitzer_herschel}

\begin{figure}[h]
\centering
        \includegraphics[width=0.5\textwidth, trim={0 0cm 0cm 0cm},clip]{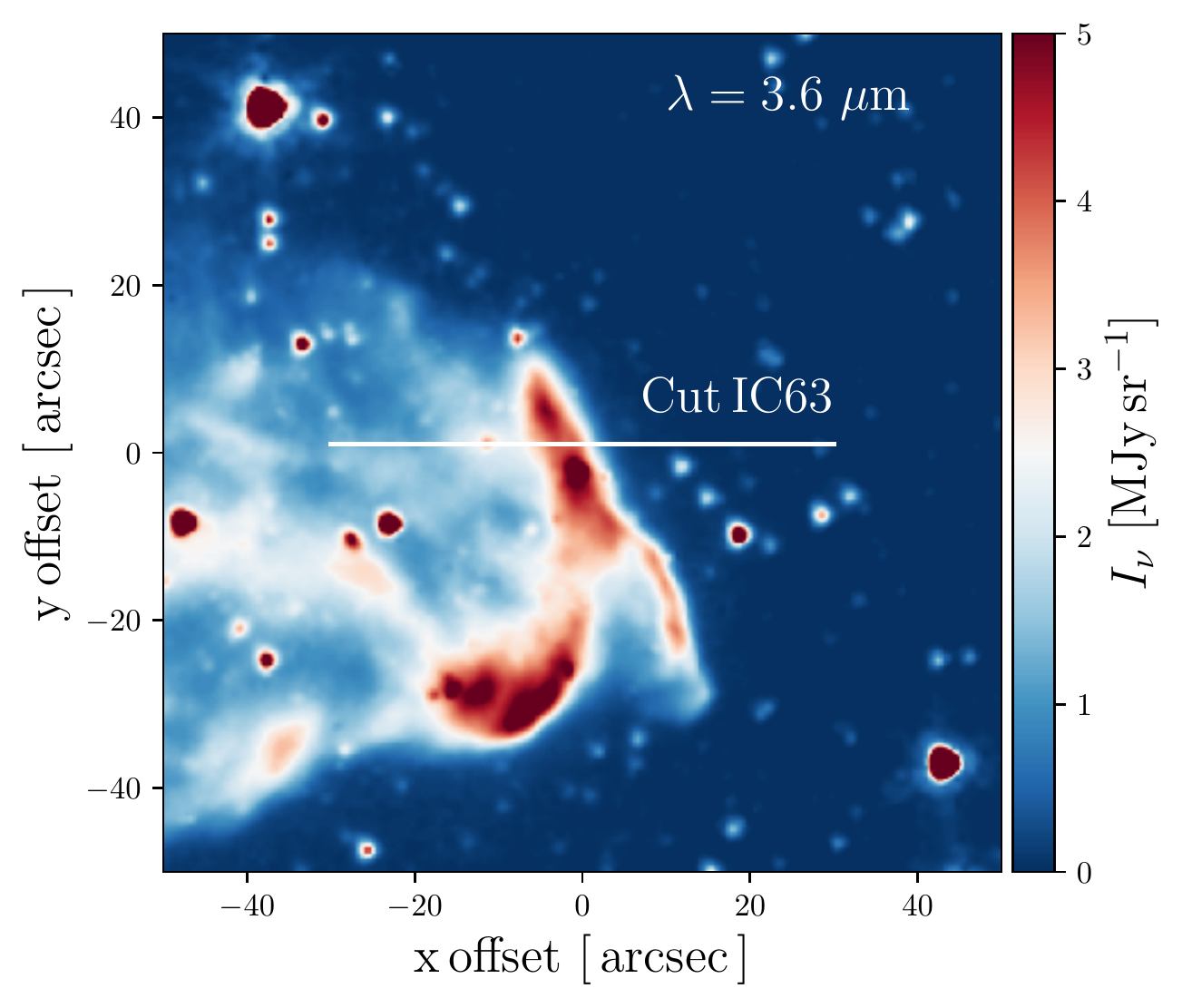}\vfill
        \includegraphics[width=0.5\textwidth, trim={0 0cm 0cm 0cm},clip]{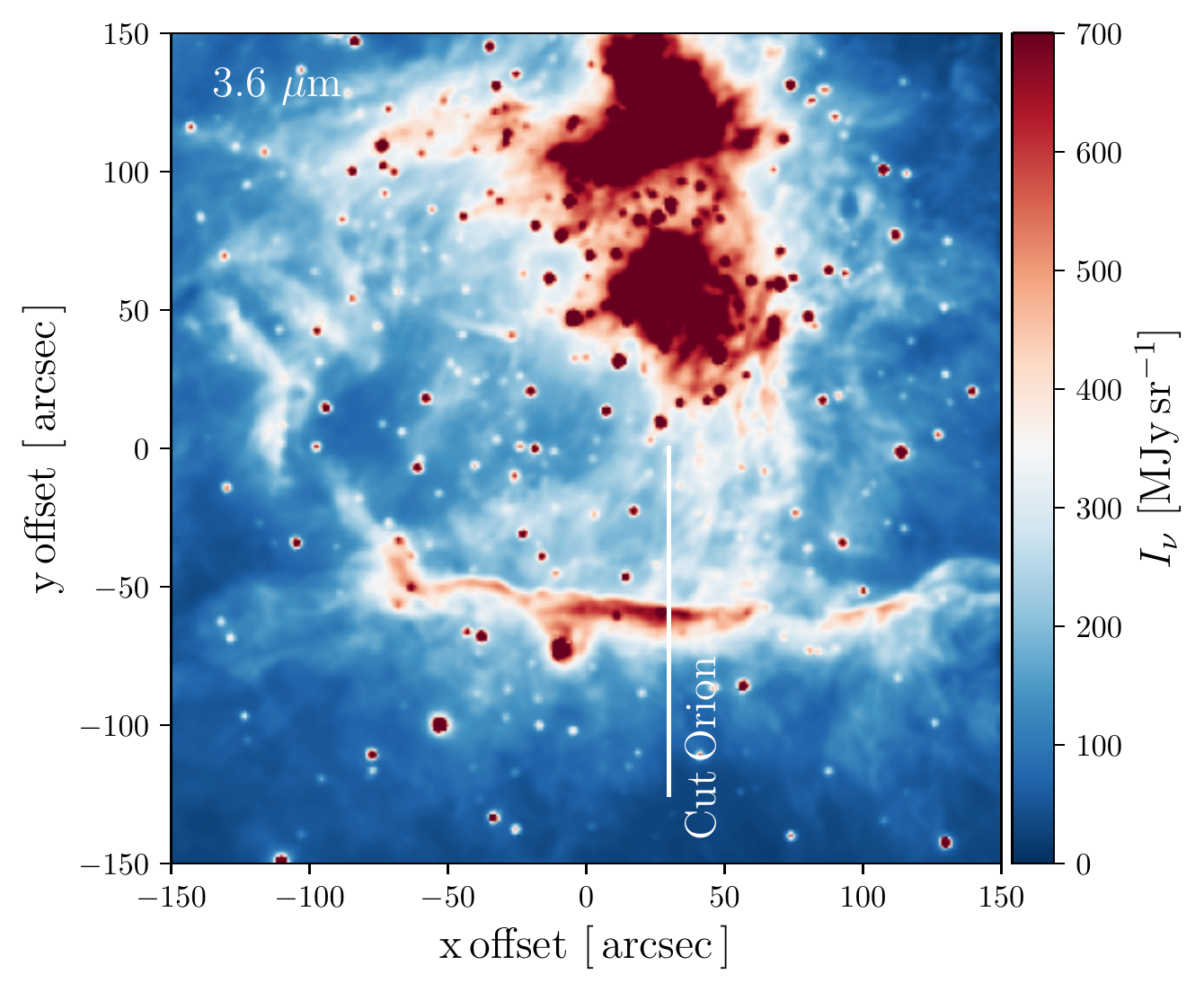}
    \caption{Selected PDRs seen with Spitzer. Top: IC63 seen at 3.6 \mum. Bottom: the Orion Bar seen at 3.6 \mum. The white solid
    lines correspond to the cuts used in our study.}
    \label{fig:IC63_orion_obs_all0}
\end{figure}

We use \spitzer~and \herschel~observations (see Fig.\,\ref{fig:IC63_orion_obs_all}) in
six photometric bands (3.6, 4.5, 5.8, 8, 24, and 70 \mum) for IC63 and in five photometric bands (3.6, 4.5, 5.8, 8, and 70 \mum) for the Orion Bar\footnote{Observations at 24 \mum~of the Orion Bar exist but are saturated.} (see Fig.\,\ref{fig:IC63_orion_obs_all}). The processing
of the \spitzer~maps is detailed in \cite{bowler_infrared_2009}. We study the observed emission profiles through a cut across both of these PDRs (see solid white lines in Fig.\,\ref{fig:IC63_orion_obs_all0}). The calibration uncertainty in the IRAC bands 
(\IRACun, \IRACdeux, \IRACtrois, and \IRACquatre) is 2 $\%$
\citep{reach_absolute_2005}, 4 $\%$ in \MIPSun~\citep{engelbracht_absolute_2007}, and 5 $\%$ in \MIPSdeux~\citep{gordon_absolute_2007}. We considered all these errors to be independent of the wavelength 
to first order. As the minor contribution of the gas in those two bands does not affect the bulk of our results, we therefore consider that the observed emission is dust emission. 

\section{Dust populations in optically thick regions}\label{sect:models_and_methods}

Photon-dominated regions are optically thick and radiative transfer is therefore required to properly model dust emission. We used the 3D radiative transfer code SOC \citep[][]{juvela_soc_2019}, together with the THEMIS dust model (described hereafter).

The Heterogeneous dust Evolution Model for Interstellar Solids\footnote{THEMIS is available here : \href{https://www.ias.u-psud.fr/themis/index.html}{https://www.ias.u-psud.fr/themis/}}
\citep[THEMIS, e.g.,][]{jones_evolution_2013, jones_global_2017} provides a description of grain properties that reflects evolutionary processes. 
In the ISM, dust evolution is mostly driven by UV photons and collisions between grains or with gas species. The efficiency of these processes, 
set by the UV flux $G_0$ and the density $n_H$, determines the structure and composition of the dust. This dust model is based on two main dust materials that are amorphous olivine-type and pyroxene-type silicates with iron and iron sulphide nano-inclusions a-Sil$_{\mathrm{Fe,\,FeS}}$, and amorphous hydrocarbons solids a-C(:H) materials \citep[see][]{jones_variations_2012, jones_variations_2012-1, jones_variations_2012-2}, which encompasses a-C:H materials that are H-rich and aliphatic-rich and a-C materials that are H-poor and aromatic-rich.

Another major contribution from THEMIS is the core-mantle description of dust grains. For 
the purpose of understanding this depiction and especially in the context of PDRs, it is important to understand the influence of UV irradiation that chiefly affects the
carbonaceous dust population. Such UV photons can photo-destroy C-H bonds and therefore, generate the creation of C=C bonds. Thus, 
the irradiation of a-C(:H) materials leads to their progressive aromatisation\footnote{The aromatisation refers to the process that transforms an aliphatic-rich carbonaceous material to 
an aromatic-rich material. In that case we will speak of photo-darkening as an aromatic rich material appears dark compared to an aliphatic rich material.}. As the typical penetration depth of a UV photon in a-C(:H) material is about 20 nm \citep[see Fig.\,15,][]{jones_variations_2012-1},
carbonaceous grains that are smaller than 20 nm are entirely photo-darkened and,
hence, aromatic-rich, these are a-C grains. Regarding larger carbonaceous grains, they
are composed of an aliphatic-rich a-C:H core surrounded by an H-poor and aromatic-rich a-C mantle assumed to be 20 nm thick, which prevents the photo-processing
of the core, thus allowing it to remain aliphatic-rich. These are the large a-C:H/a-C grains. This view provides us with a continuous description of carbonaceous grains
from the smallest, which contain aromatic cycles and are stochatiscally heated to 
the largest that are in thermal equilibrium. Regarding the silicates, it is assumed that this dust population is composed of a core of silicate, surrounded by a mantle of a-C. 

The THEMIS model for the diffuse ISM is therefore composed of three dust populations that are built upon the two materials of carbon (a-C(:H)) and silicate (a-Sil) described above. These three dust populations are defined as follows:

\begin{itemize}
    \item a-C(:H) dust population whose size distribution follows a power-law with an exponential cut-off. Since about 80 $\%$ of the mass of this population is found in grains smaller than 20 nm, and thus mostly aromatic-rich, we refer to it as \cm~grains or nano-grains indifferently in the following. 
    \item a-C(:H) dust population whose size distribution follows a log-normal law. As this population is essentially composed of a-C:H/a-C core-mantle grains (99 $\%$ of mass), we refer to it as \CM~grains, although a few a-C are included.
    \item a-Sil/a-C dust population whose size distribution follows a log-normal law.  
\end{itemize}

\begin{figure}
    \centering
    \includegraphics[width=0.5\textwidth]{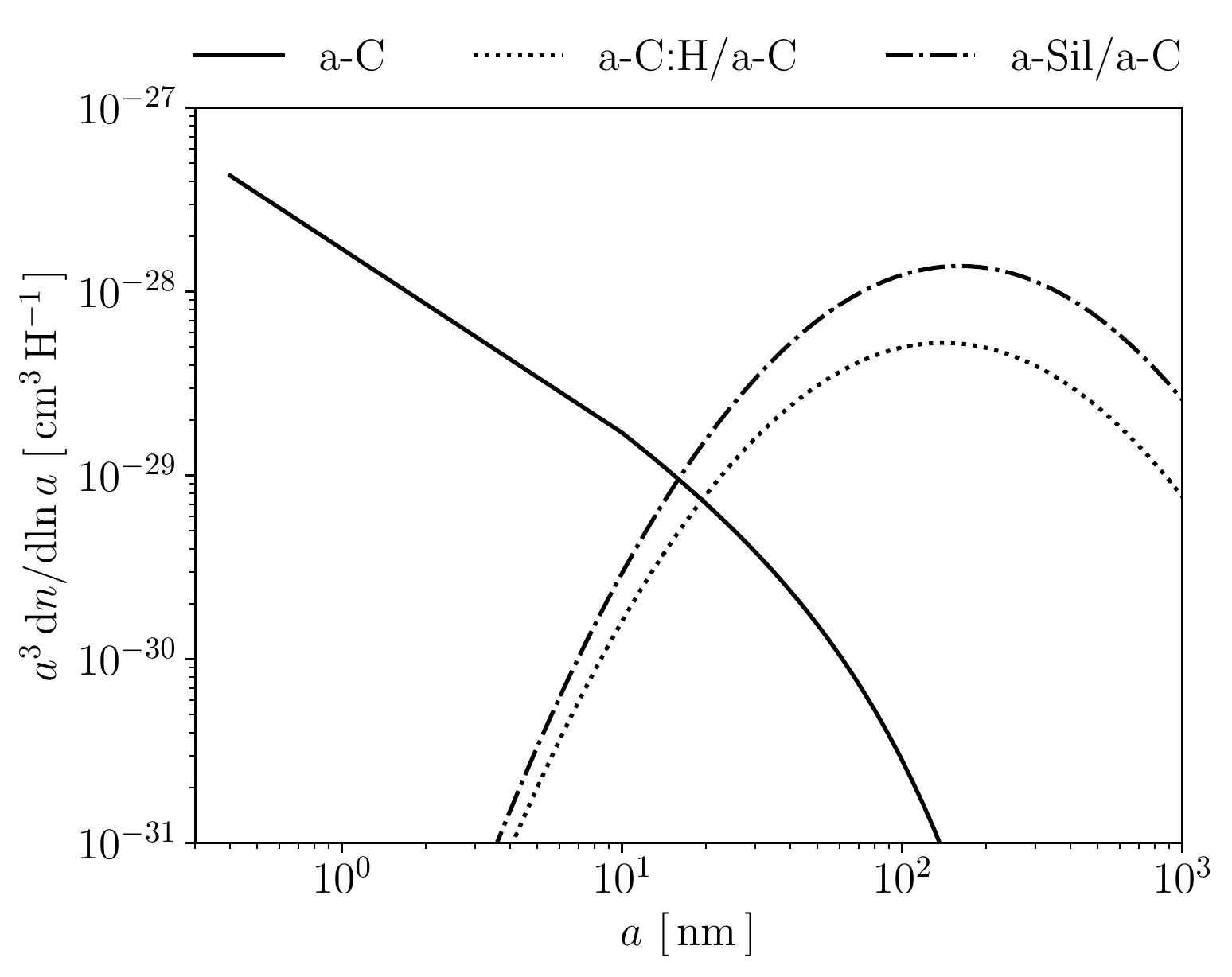}
    \caption{Dust size distributions for \cm~(solid line), \CM~(dotted line), and 
    \py~(dotted dashed line).}
    \label{fig:sdist_diffuse}
\end{figure}

The size distribution of these three populations are shown in Fig.\,\ref{fig:sdist_diffuse} and the associated parameters are detailed in Table.\,\ref{tab:parameters_size_distribution}. 

\begin{table}
\centering
\begin{tabular}{lccccccc}
    \hline\hline 
     Name & size & $\alpha$ & $a_{\mathrm{min}}$ & $a_{\mathrm{max}}$ & 
     $a_{\mathrm{c}}$ & $a_{\mathrm{t}}$ & $a_{0}$  \\
     \hline 
     \cm & p-law & 5 & 0.4 & 4900 & 10 & 50 & - \\
     \CM & log-n & - & 0.5 & 4900 & - & - & 7 \\
     \py & log-n & - & 1 & 4900 & - & - & 8 \\
     \hline
\end{tabular}
\caption{\label{tab:parameters_size_distribution} Size distribution parameters
for each dust population. p-law is a power-law with an exponential tail and log-n is 
a log-normal distribution. Sizes are given in nm.}
\end{table}

\section{Constraining the PDR structure}\label{sect:density_IC63_orion}

Radiative transfer requires information regarding the gas density profile. As we lack information about the density profile across IC63, we used a method to constrain it based only on dust emission. From now on, we take the following analytical expression to describe the density profile across these PDRs:

\begin{equation}
    \label{eq:density_profile}
    n_{\mathrm{H}}(z)=\left\{ 
\begin{array}{l l}
  n_{0} \times \left(\frac{z}{z_{0}}\right)^{\gamma}  & \quad \text{if $z<z_{0}$}\\
  n_{0} & \quad \text{if $z>z_{0}$,}\\ \end{array} \right.
\end{equation}

with $z$ as the position from the edge of the PDR, $\gamma$ the 
power-law exponent of the gas density profile, and $z_0$ the depth
beyond which constant density $n_0$ is reached. This profile has already been used in PDR studies \citep[e.g.][]{habart_density_2005, arab_evolution_2012, schirmer_dust_2020}

\subsection{IC63}\label{sect:sect:IC63_density}

No previous study has established the density profile across IC63. However, based on the studies of \cite{jansen_physical_1994} about HCO$^{+}$, CS, HCN, and H$_2$CO gas lines and \cite{jansen_physical_1995} on chemical calculations, we can set $n_0$ to $1\times 10^{5}$ H\,cm$^{-3}$. Following the studies of \cite{arab_evolution_2012} on the Orion Bar and \cite{schirmer_dust_2020} on the Horsehead, we set $\gamma = 2.5$. 

\begin{figure*}[h]
\centering
        \includegraphics[width=\textwidth, trim={0 0cm 0cm 0cm},clip]{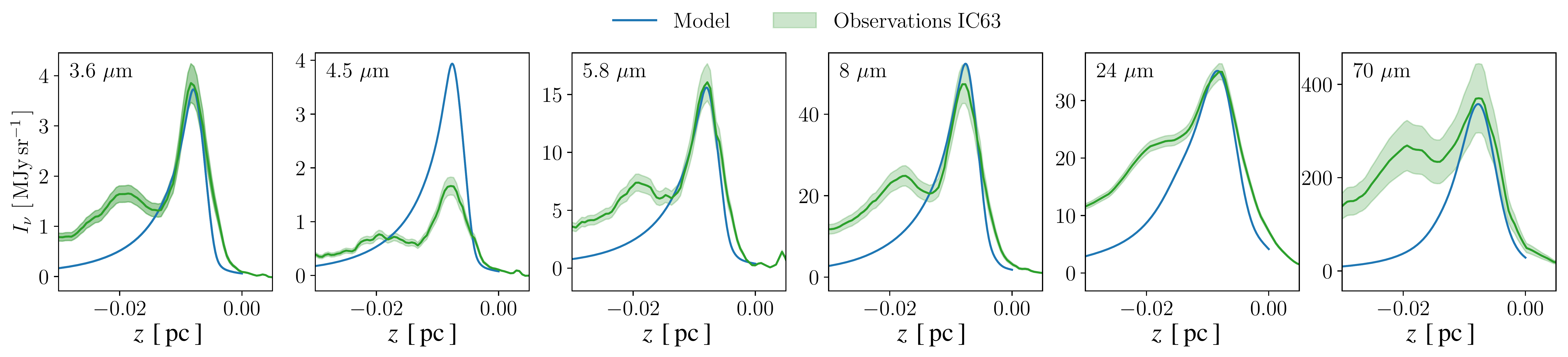}
\caption{Comparison between the observed dust emission and the modelled dust emission in IC63 using the best set of dust parameters (\abvsg~= 0.10 $\times 10^{-2}$, \aminvsg~= 0.70 nm, and $\alpha$ = -5) in six photometric bands (3.6, 4.5, 5.8, 8, 24, and 70 \mum). The dust modelled (observed) emission is shown in blue (green) line. The cut considered across IC63 is shown in Fig.\,\ref{fig:IC63_orion_obs_all}.}
    \label{fig:SOC_final_cut_2}
\end{figure*}

In Appendix \ref{appendix:width_emission}, we show that compared to variations in $z_0$, variations in the dust size distribution barely affect the width\footnote{The width of the dust emission profile can be seen in Fig.\,\ref{fig:SOC_final_cut_2}, which represent the modelled dust emission profiles using the best dust parameters and the observed dust emission profiles.} of the dust emission profiles. We therefore constrain $z_0$ through the comparison between the width of the dust modelled and observed emission profiles in the different photometric bands.

\begin{table*}[h]
    \centering
    \begin{tabular}{l|c|c|c|c|c|c}
        \hline
        \hline
        & 3.6 \mum & 4.5 \mum & 5.8 \mum & 8 \mum & 24 \mum & 70 \mum \\
         \hline
        IC63 $(\Delta_{i,\,\mathrm{obs}})$ & 0.0026 & 0.0028 & 0.0027 & 0.0030 & 0.0045 & 0.0038 \\
        Orion Bar $(\Delta_{i,\,\mathrm{obs}})$ & 0.010 & 0.010 & 0.010 & 0.010 & - & 0.0137  \\
        \hline
        IC63 $(I_{\mathrm{obs,\,max}})$ & 3.8 & 1.7 & 16.1 & 47.4 & 35.0 & 370.3 \\
        Orion Bar $(I_{\mathrm{obs,\,max}})$ & 650 & 399 & 815 & 6645 & - & 3.23 $\times 10^{5}$ \\
        \hline
    \end{tabular}
    \caption{FWHM ($\Delta_{i,\,\mathrm{obs}}$, expressed in pc) and maximum intensity ($I_{\mathrm{obs},\,\mathrm{max}}$, expressed in MJy\,sr$^{-1}$) of the observed dust emission profiles across IC63 and the Orion Bar (cuts across IC63 and the Orion Bar are showed in Fig.\,\ref{fig:IC63_orion_obs_all}) in six photometric bands.}
    \label{tab:obs_FWHM}
\end{table*}
 
We computed dust emission across IC63 using THEMIS for $z_0$ that varies from 0.001 pc to 0.020 pc on a linear grid of 100 points. We then compared the full width at half maximum (FWHM) of the observed and modelled dust emission profiles by minimising the following $\chi_{\mathrm{wid}}^{2}$:

\begin{equation}
    \label{eq:chi2}
    \chi_{\mathrm{wid}}^{2} = \sum_{i\;\in\;\mathrm{filters}} \left(\Delta_{i,\,\mathrm{mod}}-\Delta_{i,\,\mathrm{obs}} \right)^{2} \;,  
\end{equation}

where $\Delta_{i,\,\mathrm{mod}}$ $(\Delta_{i,\,\mathrm{obs}})$ is the FWHM of the modelled (observed) dust emission profiles in the $i$-th band. The FWHM of the observed dust emission profiles from 3.6 \mum~to 70 \mum~for both IC63 and the Orion Bar can be found in Table\,\ref{tab:obs_FWHM}. 

\begin{figure*}[h]
\centering
\includegraphics[width=0.5\textwidth, trim={0 0cm 0cm 0cm},clip]{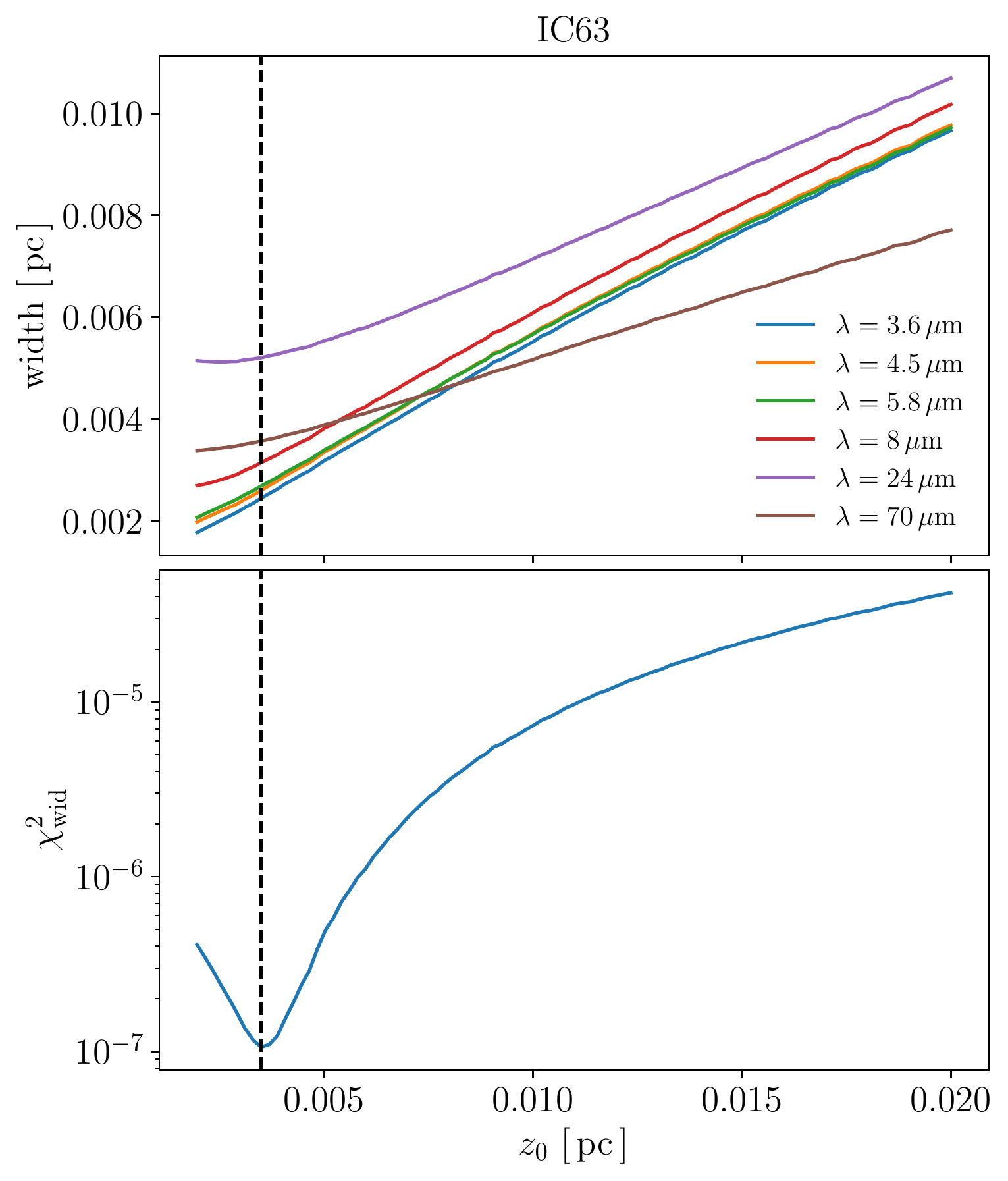}\hfill
\includegraphics[width=0.5\textwidth, trim={0 0cm 0cm 0cm},clip]{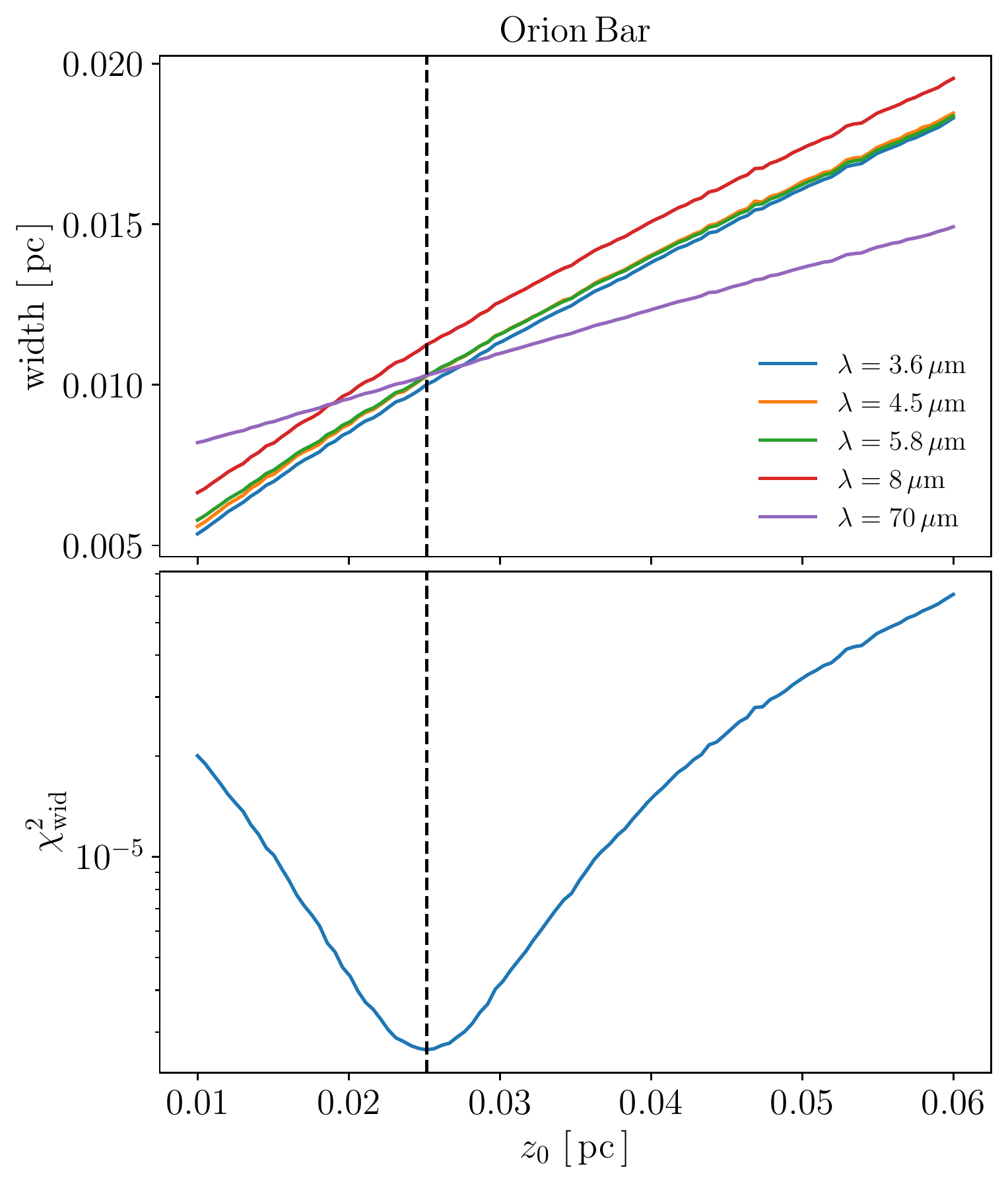}
    \caption{$\chi_{\mathrm{wid}}^2$ minimisation for IC63 and the Orion Bar. Top: FWHM of the observed dust emission profiles in different photometric bands as a function of $z_0$ (left for IC63 and right for the Orion Bar). Bottom: $\chi_{\mathrm{wid}}^2$ as a function of $z_0$ (left for IC63 and right for the Orion Bar). The vertical black dashed lines corresponds to the minimum value of $\chi_{\mathrm{wid}}^2$. Radiative transfer has been done using diffuse ISM-like dust with an incident radiation field corresponding to a blackbody at 25\,000 K (38\,000 K) with \G~= 1100 (\G~= 2.6 $\times$ 10$^{4}$) for IC63 (Orion Bar).}
    \label{fig:z0_width_chi2}
\end{figure*}

We show the results of this minimisation in Fig.\,\ref{fig:z0_width_chi2} (left panels). Regardless of the band, the FWHM increases with $z_0$ (see Fig.\,\ref{fig:z0_width_chi2}, top left panel) which is expected as the larger $z_0$ the smoother the density profile slope hence the radiation can penetrate deeper in the cloud to heat dust grains, thus increasing the FWHM. Regarding the $\chi^2$ minimisation (see Fig.\,\ref{fig:z0_width_chi2}, bottom left panel), there is a strong minimum for $z_0 = 0.004$ pc which we adopt in the following. To summarise, the parameters that describe the density profile across IC63 are: 

\begin{equation}
        \label{eq:density_parameters}
    n_{0} = 1 \times 10^{5}\,\mathrm{H\,cm^{-3}} \quad ; \quad 
    z_{0} = 0.004\,\mathrm{pc} \quad ; \quad 
    \gamma = 2.5.
\end{equation}

\subsection{Orion Bar}\label{sect:sect:Orion_density}

Conversely to IC63, many studies have been carried out with the aim of exploring the Orion Bar (see Sect.\,\ref{sect:sect:orion}). In our study, we use the density profile from \cite{arab_evolution_2012}. Whilst it provides a value of $z_0$ - in addition to $n_0$ and $\gamma$-, we choose to constrain it using the same procedure as for IC63. 

We show the results of the $\chi^2$ minimisation in Fig.\,\ref{fig:z0_width_chi2} (right panels). There is a clear minimum for $z_0$ = 0.025 pc. One can note that this value is nearly twice as small as in the study of \cite{arab_evolution_2012}. This discrepancy is probably due to the radiative transfer modelling. Indeed, \cite{arab_evolution_2012} used a 1D radiative transfer code whereas we use a 3D radiative transfer code. In their modelling, it is assumed that 88 $\%$ of the photons are forward scattered (i.e. 12 $\%$ are back scattered); however, in a 3D configuration, a more extensive treatment of the scattering will lead to less forward scattering photons. Nevertheless, those differences barely affect the dust properties that are obtained in both studies. To summarise, the parameters that describe the density profile across the Orion Bar are the following:

\begin{equation}
        \label{eq:density_parameters_orion_bar}
    n_{0} = 1.5 \times 10^{5}\,\mathrm{H\,cm^{-3}} \quad ; \quad 
    z_{0} = 0.025\,\mathrm{pc} \quad ; \quad 
    \gamma = 2.5.
\end{equation}

We note that $z_0$ is six times larger in the Orion Bar than in IC63. However, we cannot conclude that the density profile in situ is steeper in IC63 as it could also be due to spatial resolution effects. 

\section{Constraining dust properties}\label{sect:dust_prop}

The approaches pursued to constrain the dust properties in IC63 and the Orion bar are slightly different due to the specificity of those two objects. In this section, we present those approaches and the constraints on the nano-grain properties obtained with the radiative transfer modelling. A summary of the results can be found in Table\,\ref{tab:resume2}.

\subsection{Two objects, two approaches}\label{sect:approaches}

Before going further, it is necessary to present the parameters associated with the nano-grain properties we aim to constrain. Three parameters associated with their size distribution are varied: 1) the abundance, that is, the a-C mass to gas ratio, \abvsg; 2) the minimum size, \aminvsg; and 3) the slope of the power-law size distribution, $\alpha$. The influence of variations in these parameters on both the dust size distribution and the associated spectra in the optically thin limit are shown in Paper\,I (Fig.\,4, first line for the dust size distributions and second line for the associated spectra). Another important parameter is the length of the PDR along the line of sight, \lpdr. As this parameter does not affect the shape of the dust spectrum (see Sect.\,4.2 in Paper\,I), \lpdr~can be adjusted after the fact. 

A simplified approach would be to explore the 3D space defined by \abvsg, \aminvsg, and $\alpha$ and then computing the radiative transfer for each set of parameters. However, if we assume a subset of $N$ values for each parameter, the computation time would be proportional to $N^{3}$, which is time-consuming. In our case, as discussed below, it is possible to divide the exploration of the 3D space into a 2D exploration (\aminvsg, $\alpha$) and then follow up with a 1D (\abvsg) exploration\footnote{The computation time is therefore proportional to $(N^{2}+N)$ instead of $N^{3}$. The time saved using the 2D+1D exploration instead of the 3D exploration is proportional to $\left(1+N\right)^{-1}\rightarrow 1/N$ for $N>>1$. For a basic grid of $10\times 10\times 10$ parameters, this leads to total computation time almost ten times lower when exploring the 2D+1D spaces as compared to that of the 3D space.}.

In the case of the Orion Bar, the observations consist of four photometric bands in the mid-IR (MIR, i.e. 3.6, 4.5, 5.8, and 8 \mum) and one in the NIR/FIR (70 \mum). Variations in \aminvsg~and $\alpha$ change the dust spectral shape (i.e. ratios between those four bands evolve non-linearly) in the MIR but barely affect the FIR dust emission (see Fig.\,4 in Paper\,I). On the contrary, variations in \abvsg~do not affect the spectral shape in the MIR but, instead, the MIR-to-FIR dust emission ratio\footnote{See Sect.\,4.3 of paper\,I for details.} only. In this specific case, we therefore adjust the shape of the MIR dust spectrum by exploring the 2D space (\aminvsg, $\alpha$) and then we adjust the MIR to FIR dust emission ratio by varying \abvsg. To this end, we define:

\begin{equation}
    \label{eq:chi2_MIR}
    \chi_{\mathrm{MIR}}^{2} = \sum_{i\,\in\,\mathrm{filters(MIR)}} \left(\frac{X_{i}-\mu_{\mathrm{MIR}}}{\sigma_{i}} \right)^{2} \;,  
\end{equation}
and
\begin{equation}
    \label{eq:chi2_MIR1}
    \chi_{\mathrm{tot}}^{2} = \sum_{i\,\in\,\mathrm{filters}} \left(\frac{X_{i}-\mu_{\mathrm{tot}}}{\sigma_{i}} \right)^{2} \;,  
\end{equation}
with
\begin{equation}
    \label{eq:chi2_1}
    X_{i} = \frac{I_{\mathrm{obs,max}}(i)}{I_{\mathrm{mod,max}}(i)} \quad ; \quad 
    \sigma_{i} = r_{\mathrm{obs}}(i)\,X_{i},
\end{equation}
and
\begin{equation}
    \label{eq:chi2_2}
    \mu_{\mathrm{MIR}} = \left< X_{i} \right>_{i\,\in\,\mathrm{filters(MIR)}} \quad ; \quad \mu_{\mathrm{tot}} = \left< X_{i} \right>_{i\,\in\,\mathrm{filters}},
\end{equation}

where $r_{\mathrm{obs}}$ is the relative error for each band and defined in
Sect.\,\ref{sect:sect:spitzer_herschel} and $I_{{\mathrm{obs,\,max}}}(i) = \mathrm{max}\left(I_{\mathrm{obs},\,i}(z)\right)$
with $I_{\mathrm{obs},\,i}(z)$, 
the dust observed in the $i$-th band at the position $z$ 
along the cut. \chiMIR~only takes into account the MIR bands and is therefore sensitive to variations in \aminvsg~and $\alpha$. Conversely, \chiMIR~does not depend on \abvsg. We therefore adjust the shape of the MIR dust spectrum by minimising \chiMIR~in the 2D space (\aminvsg, $\alpha$) and then we adjust the overall MIR to FIR ratio by minimising \chiTOT.

In the case of IC63, it is not possible to apply the same method as variations in \abvsg~lead to a non-linear evolution of the ratio between the 24 \mum~band with each of the four bands in the MIR. We therefore have to minimise \chiTOT~in the 3D space (\abvsg, \aminvsg, $\alpha$).

\subsection{IC63}\label{sect:dust_prop_IC63}

We study the \chiTOT~distribution in the 3D space (\abvsg, \aminvsg, and $\alpha$), defined as follows:
\begin{enumerate}
    \item \abvsg~varies from 0.05 $\times$ $10^{-2}$ to 0.017 $\times$ $10^{-2}$ on a 10-step linear grid.
    \item \aminvsg~varies from 0.4 nm to 0.9 nm on a 10-point linear grid.
    \item $\alpha$ varies from -4.5 to -5.6 in steps of 0.1.
\end{enumerate}

\begin{figure}[h]
\centering
\includegraphics[width=0.5\textwidth, trim={0 0cm 0cm 0cm},clip]{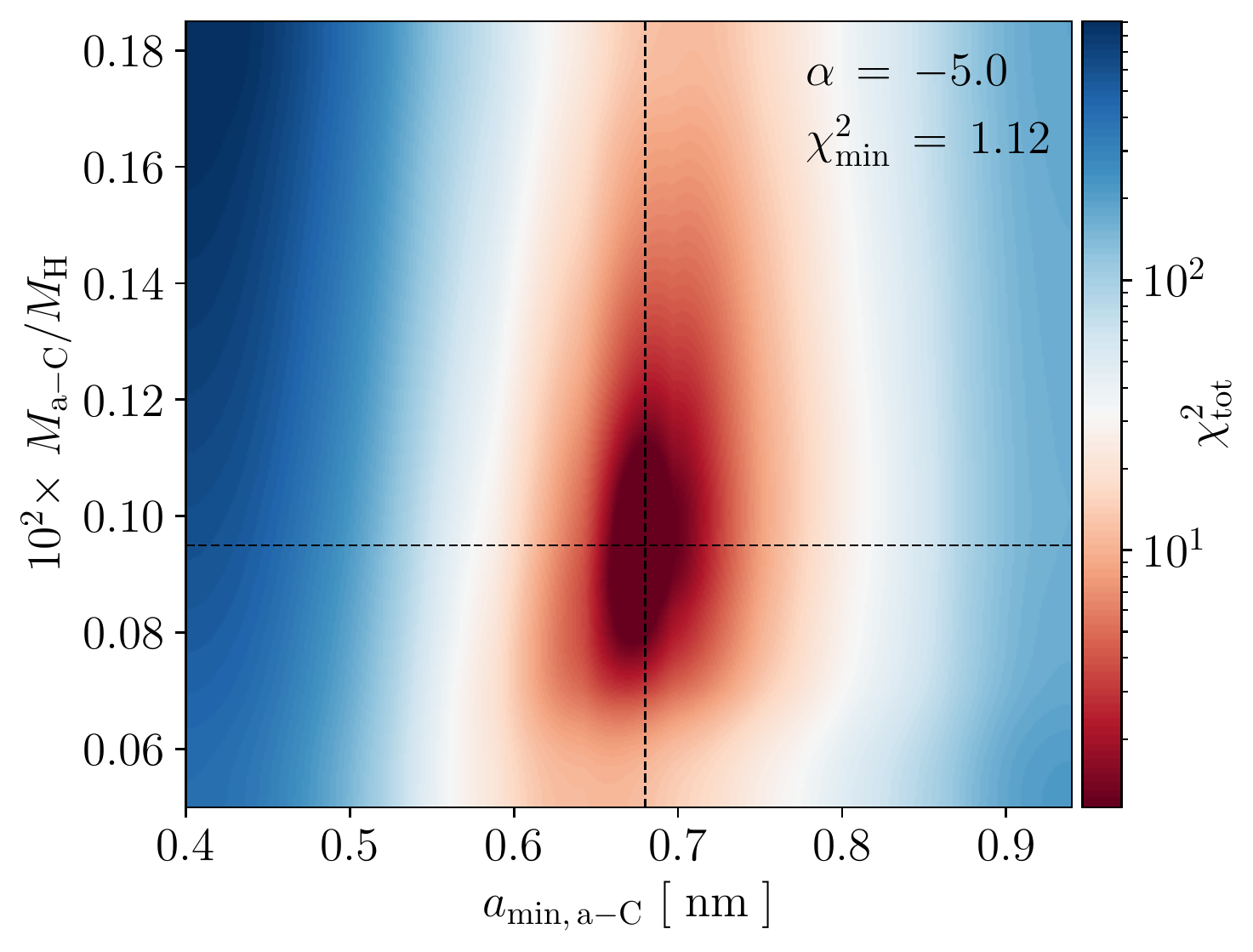}
    \caption{\chiTOT~in the 2D space (\abvsg~and \aminvsg) for $\alpha=-5$ in the IC63 model.}
    \label{fig:2D}
\end{figure}

\begin{figure}[h]
\centering
\includegraphics[width=0.5\textwidth, trim={0 0cm 0cm 0cm},clip]{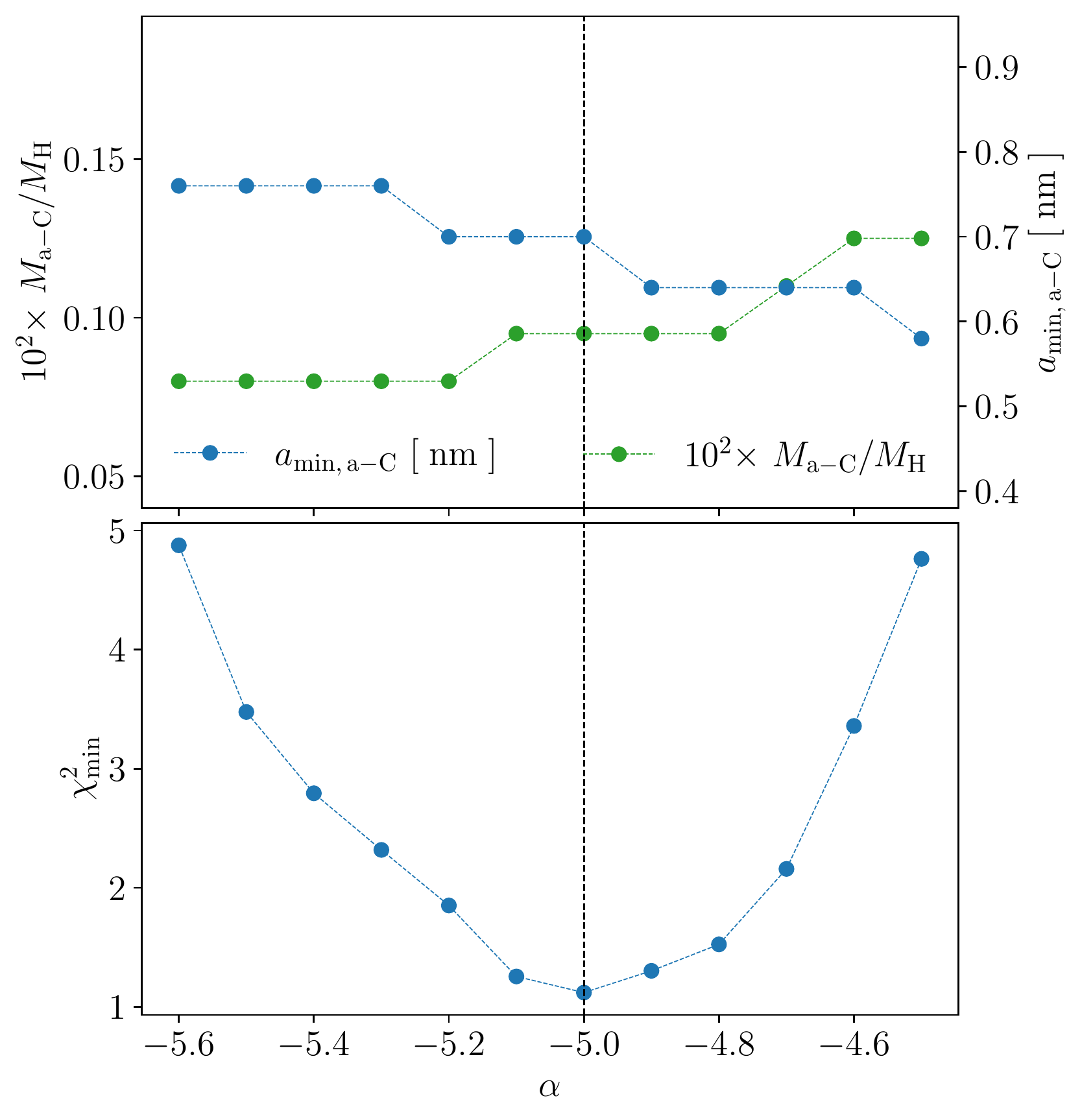}
    \caption{Top: \abvsg~and \aminvsg~associated with the minimum \chiTOT~in the 2D space (\abvsg, \aminvsg) as a function of $\alpha$ (see Fig.\,\ref{fig:SOC_grid_cut_1_n0_1e5}). Bottom: Minimum value of \chiTOT~in the 2D space (\abvsg, \aminvsg) as a function of $\alpha$.}
    \label{fig:alpha_chi2}
\end{figure}

Figure\,\ref{fig:2D} shows \chiTOT~in the 2D space (\abvsg, \aminvsg) for $\alpha=-5, $  while Fig.\,\ref{fig:alpha_chi2} shows the minimum value of \chiTOT~in the 2D space (\abvsg, \aminvsg) as a function of $\alpha$. We also show in Fig.\,\ref{fig:SOC_grid_cut_1_n0_1e5}, \chiTOT~in the 2D space (\abvsg, \aminvsg) for different values of $\alpha$.  

The \aminvsg~value associated with the minimum of \chiTOT, increases with a decrease in $\alpha$. Indeed, a decrease in $\alpha$ leads to a enhancement of the nano-grains that are responsible for the emission in the NIR/MIR. This is counterbalanced by an increase in \aminvsg~, since such a variation of \aminvsg~leads to a decrease of the smallest of the nano-grains. 

For each $\alpha$ value tested, there is a different minimum in the 2D space (\abvsg, \aminvsg). This means that there are no degeneracies and that there is an absolute minimum in the 3D space (\abvsg, \aminvsg, and $\alpha$). To locate the absolute minimum, we show the minimum value of \chiTOT~in the 2D space (\abvsg, \aminvsg) as a function of $\alpha$ in Fig.\,\ref{fig:alpha_chi2}, as well as the values of \abvsg~and \aminvsg~associated with these minima. We find a unique minimum for $\alpha=-5$, which gives \abvsg~= $0.10\times10^{-2}$ and \aminvsg~= 0.7 nm.

We show in Fig.\,\ref{fig:SOC_final_cut_2}, the dust modelled and observed emissions across IC63 for the best-fit parameters. The width of the dust emission profiles reproduces generally well the one of the dust observed profiles, which reinforces our hypothesis where the width of the dust emission profiles barely depends on the dust properties and solely depends on the density profile. Except at 4.5 \mum, the dust emission maxima are well reproduced. The discrepancy at 4.5 \mum~worsens as this band has been deliberately excluded from the \chiTOT~minimisation because we are not able to simultaneously fit the observations in all of the six photometric bands. This discrepancy has already been observed and discussed in more details in Sect.\,6.1 of Paper\,I for the Horsehead Nebula.

\subsection{Orion Bar}\label{sect:dust_prop_orion}

\subsubsection{Contraining \aminvsg~and $\alpha$}\label{sect:a_min_alpha}

We study the \chiMIR~distribution in the 2D space (\aminvsg~and $\alpha$), defined as follows:
\begin{enumerate}
    \item \aminvsg~varies from 0.35 nm to 0.9 nm on a 40-point linear grid;
    \item $\alpha$ varies from -3 to -13 on a 40-point linear grid.
\end{enumerate}

Figure \ref{fig:orion_chi2_amin_alpha} (left panel) shows the result. We observe a clear degeneracy between \aminvsg~and $\alpha$, the first increasing when the second decreases. This is expected as a decrease in $\alpha$ implies an increase in the NIR/MIR dust emission, which is counterbalanced by an increase in \aminvsg. We also see that the \chiMIR~is lower than eight in the degeneracy, which means that the dust MIR spectrum is well reproduced. We choose five couples of \aminvsg~and $\alpha$ (see the symbols in Fig.\,\ref{fig:orion_chi2_amin_alpha}, left panel) to cover the whole degeneracy and then constrain \abvsg. Finally, it is important to note that despite the degeneracy $\alpha \leq 6$, which is linked to the fragmentation of large grains into nano-grains (see Sect.\,\ref{sect:sect:fragmentation}) and \aminvsg\,$\leq 0.8$ nm, which is linked to the photo-destruction of nano-grains (see Sect.\,\ref{sect:sect:photodestruction_a-C}).

\subsubsection{Constrain \abvsg}\label{sect:ab_vsg}

We studied the \chiTOT~distribution in the 1D space (\abvsg), where \abvsg~varies from $1\times 10^{-5}$ to $0.17\times 10^{-2}$ on a logarithmic scale composed of 25 points. According to the previous section, we take five different couples of \aminvsg~and $\alpha$ (1: 0.35 nm/-5.96, 2: 0.51 nm/-5.96, 3: 0.63 nm/-7.44, 4: 0.71 nm/-8.92, 5: 0.77 nm/-12.25) that minimise \chiMIR.

\begin{figure*}[h]
\centering
        \includegraphics[width=0.5\textwidth, trim={0 0cm 0cm 0cm},clip]{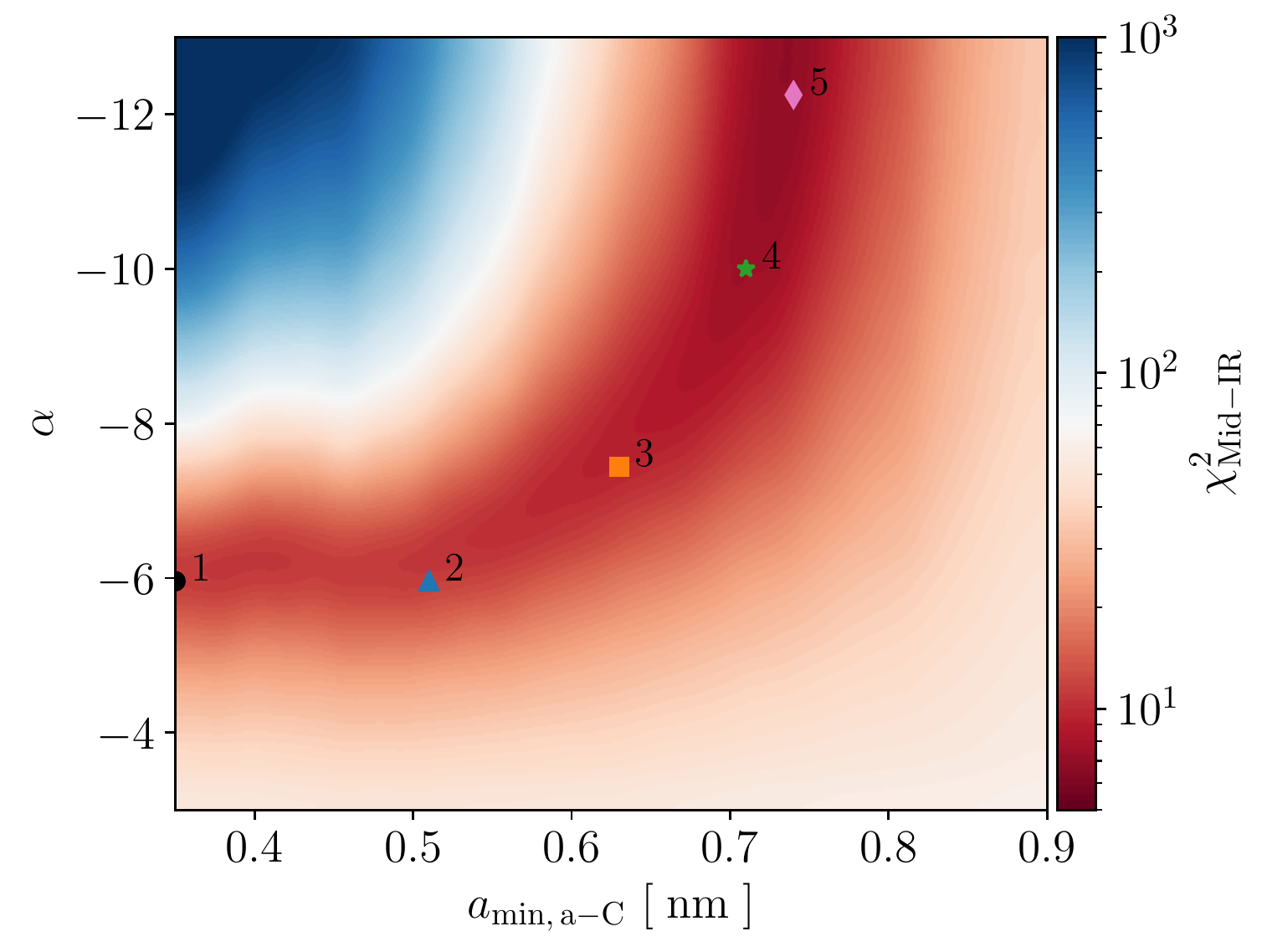}\hfill 
        \includegraphics[width=0.5\textwidth, trim={0 0cm 0cm 0cm},clip]{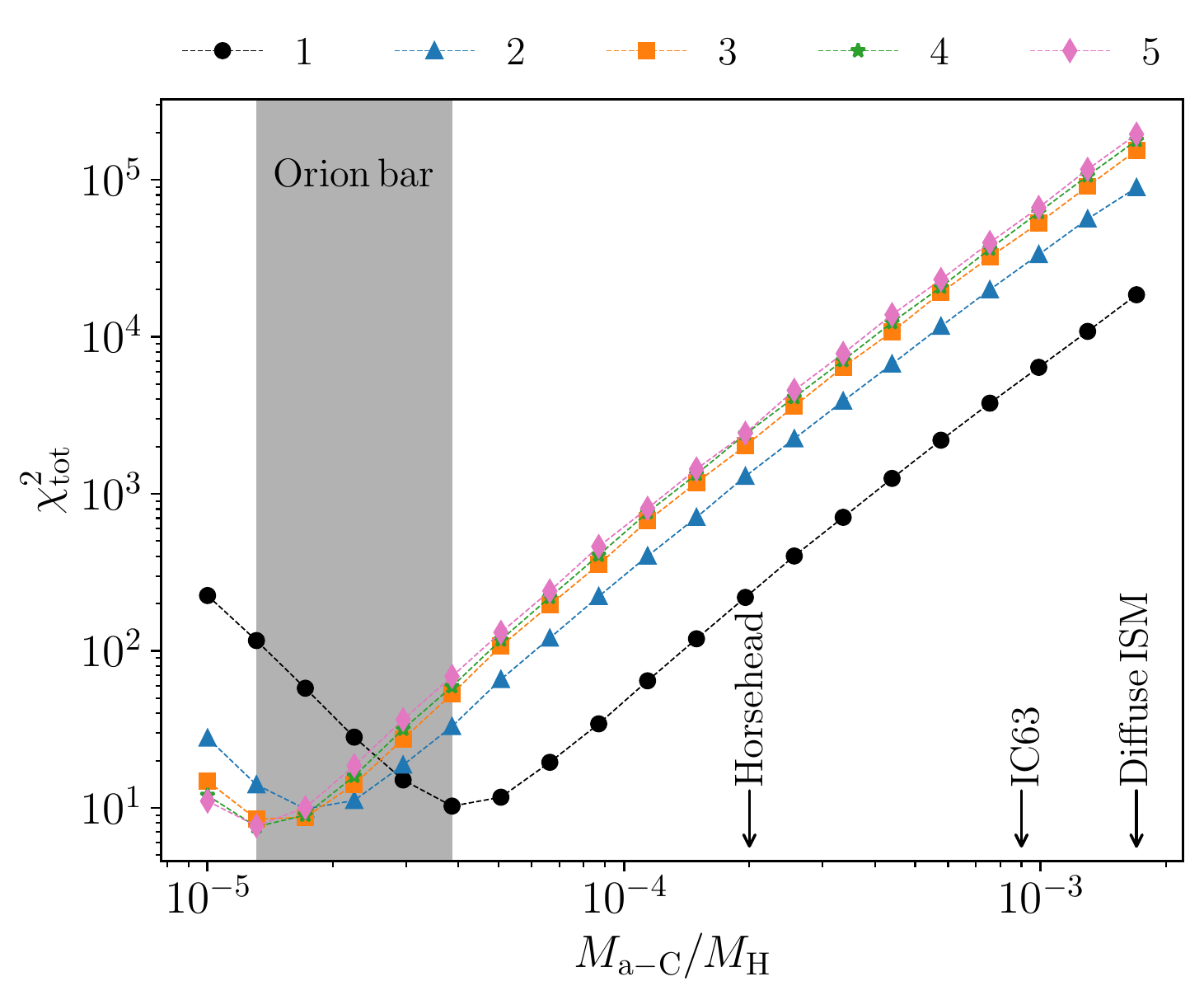}
\caption{$\chi^2$ minimisation for the Orion Bar. Left: \chiMIR~in the 2D space (\aminvsg~and $\alpha$). Right: \chiTOT~as a function of \abvsg~for five different couple \aminvsg/$\alpha$, that are represented by the five symbols on the left figure. 1 (black dot): 0.35 nm/-5.96, 2 (blue triangle): 0.51 nm/-5.96, 3 (orange square): 0.63 nm/-7.44, 4 (green star): 0.71 nm/-8.92, 5 (pink diamond): 0.77 nm/-12.25. We also indicate with black arrows the \abvsg~obtained for the Horsehead (see Paper\,I), IC63 (see Sect.\,\ref{sect:dust_prop_IC63}), and the diffuse ISM \citep{jones_evolution_2013, jones_global_2017}. The vertical grey stripe corresponds to the range of \abvsg~that minimise \chiTOT~in the Orion Bar.}
    \label{fig:orion_chi2_amin_alpha}
\end{figure*}

\begin{figure*}[h]
\centering
        \includegraphics[width=\textwidth, trim={0 0cm 0cm 0cm},clip]{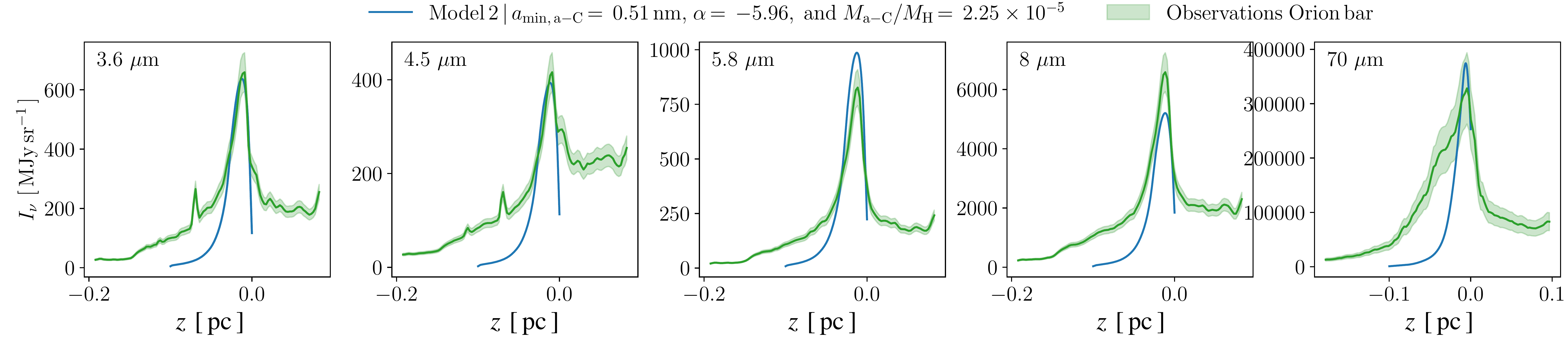}
\caption{Comparison between the observed dust emission and the modelled dust emission using the best set of dust parameters, model 2 in that case (see Fig.\,\ref{fig:orion_chi2_amin_alpha}) in five photometric bands (3.6, 4.5, 5.8, 8, and 70 \mum). The dust observed emission is shown in green line. The cut considered across the Orion Bar is shown in Fig.\,\ref{fig:IC63_orion_obs_all}.}
    \label{fig:SOC_final_1_cut_orion}
\end{figure*}

Figure \ref{fig:orion_chi2_amin_alpha} (right panel) shows the results. There is a minimum value for each couple \aminvsg/$\alpha$, which provides a range of \abvsg~values from $\sim 1\times 10^{-5}$ to $\sim 4\times 10^{-5}$. To summarise, while there is a degeneracy between the best \aminvsg~and $\alpha$ values, we are able to provide a constraint on \abvsg. We show in Fig.\,\ref{fig:SOC_final_1_cut_orion} the comparison between the dust observed emission and the dust modelled emission for one of the five different set of best parameters. The same figures for the four other models can be found in Fig.\,\ref{fig:IC63_orion_obs_all}. 

Conversely to the Horsehead and IC63, we are able to fit simultaneously the four IRAC bands, including the band at 4.5 \mum. This is interesting as albeit the bands at 3.6, 5.8, and 8 microns cover aromatic and aliphatic features, the 4.5 band almost only covers the dust continuum (see Fig.\,3 in Paper\,I). This means that contrary to the Horsehead and IC63 cases, the band to continuum ratio is well reproduced in the Orion Bar (as well as in protoplanetary disks, see \cite{bouteraon_carbonaceous_2019} for instance). Therefore, THEMIS in highly irradiated regions (Orion Bar, protoplanetary disks, etc.) seems to work well, however, it needs some further developments in low to moderate PDRs. Spectral observations that will be accessible with the JWST will most likely bring more conclusive constraints.

\begin{table*}[h]
    \centering
    \begin{tabular}{lccccccccc}
        \hline
        \hline
        PDR & $G_0$ & $T_{\mathrm{eff}}$ & $D$ & \abvsg & \aminvsg & $\alpha$ & $n_0$ & $z_0$ & $\gamma$ \\
        & & [\,K\,] & [\,pc\,] &  & [\,nm\,] & & [\,H\,cm$^{-3}$\,] & [\,pc\,] & \\
         \hline
          IC63 & 1100 & 25000 & 1 & $(0.1\pm 0.01)\times 10^{-2}$ & $0.70\pm 0.01$ & $-5\pm 0.1$ & $1 \times 10^{5}$ & 0.004 & 2.5 \\
          Horsehead & 100 & 35000 & 3.5 & $(0.02\pm 0.01)\times 10^{-2}$ & $0.77\pm 0.03$ & $-6\pm 0.5$ & $2 \times 10^{5}$ & 0.06 & 2.5 \\
          Orion Bar & 26000 & 38000 & 0.25 & $(0.0025\pm 0.0015)\times 10^{-2}$ & $\leq 0.8$ & $\leq -5.5 $ & $1.5 \times 10^{5}$ & 0.025 & 2.5 \\
        \hline
        Diffuse ISM & 1 & - & - & $0.17\times 10^{-2}$ & 0.4 & -5 & - & - & - \\
        \hline
    \end{tabular}
    \caption{Summary of the physical parameters defining the irradiation ($G_0$, $T_{\mathrm{eff}}$) and the density profiles ($n_0$, $z_0$, $\gamma$) as well as the results of the dust property modelling (\abvsg, \aminvsg, $\alpha$).}
    \label{tab:resume2}
\end{table*}

\section{Discussion}\label{sect:discussion}

Here, we present the constraints on dust properties we obtain in IC63 and the Orion Bar (Sect.\,\ref{sect:sect:main_results}). We also summarise the results in the Horsehead from Paper I. Thence we present the mechanisms that are most likely at the origin of the nano-grain evolution in PDRs, whether for their destruction (Sect.\,\ref{sect:sect:photodestruction_a-C}) or formation (Sect.\,\ref{sect:sect:fragmentation}). We finally discuss these results and propose a scenario for dust evolution in PDRs (Sect.\,\ref{sect:sect:dust_evolution}).

\subsection{Main results}\label{sect:sect:main_results}

We use the 3D radiative transfer code SOC together with THEMIS to model dust emission in IC63 and the Orion Bar. We find a good agreement between our modelled dust emission and the observations of \spitzer~and \herschel~in IC63 (see Fig.\,\ref{fig:IC63_orion_obs_all}) and the Orion Bar (see Fig.\,\ref{fig:SOC_final_cut_orion}). 

We first constrain the depth threshold above which the density profile reaches a maximum (i.e. $z_0$) only based on the width of the dust modelled emission, which does not depend on the dust properties. We then assess dust properties in IC63 and in the Orion Bar. The main results are as follows: 

\begin{enumerate}
    \item Whether in IC63 or in the Orion Bar, the nano-grain dust-to-gas mass ratio, \abvsg, is lower than in the diffuse ISM. In IC63, \abvsg~is roughly twice as low as in the diffuse ISM. In the Orion Bar, \abvsg~is 60-100 times lower than in the diffuse ISM. The uncertainty on this last value is due to the degeneracy between \aminvsg~and $\alpha$. 
    \item The nano-grain minimum size, \aminvsg, in IC63 is 1.75 times larger than in the diffuse ISM. In the Orion Bar, the degeneracy between \aminvsg~and $\alpha$ does not allow us to draw a conclusion on the nano-grain minimum size.
    \item The power-law exponent of the nano-grain size distribution, $\alpha$, is the same in IC63 and in the diffuse ISM. Regarding the Orion Bar, $\alpha$ is at least 1.2 times lower than in the diffuse ISM.
\end{enumerate}

In IC63, it is possible to simultaneously fit the observations in all the photometric bands only if we remove the one at 4.5 \mum. This problem has already been encountered in the Horsehead Nebula (see Paper I) and explained therein. Regarding the Orion Bar, we do not have this problem because, whether we use this band or not, we still have a degeneracy between \aminvsg~and $\alpha$. In addition, this does not affect our results on the dust properties. These results confirm what was found for the Horsehead, namely, an increase in the nano-grain minimum size together with a decrease in the nano-grain dust-to-gas mass. For illustrative purposes, the spectra obtained for the Orion Bar are used in \cite{berne_pdrs4all_2022} and compared to spectral observations at several representative PDR positions. We also show, in Appendix\,\ref{appendix:near2farIRratio}, a study of the NIR-to-FIR emission ratio across the Orion bar that is purely observation-based and therefore not model-dependent, and which is in agreement with a decrease in the nano-grain abundance when one moves away from the illuminating star.

\subsection{Photo-destruction of nano-grains}\label{sect:sect:photodestruction_a-C}

In addition to being stochastically heated, nano-grains can be destroyed by energetic photons \citep[e.g.][]{alata_vacuum_2014, alata_vacuum_2015}. As a-C(:H) nano-grains are composed of a mix of different molecular domains, that is, aromatic domains connected by aliphatic (C-C) and olefinic (C=C) bridges, their photo-destruction is triggered by the photo-dissociation of aliphatic and olefinic bonds. From there, there are at least three processes that can lead to the photo-destruction of nano grains: direct dissociation, thermal (vibrational) dissociation, and Coulomb explosion \citep[see][and references therein]{montillaud_evolution_2013}.


However, the probability that an absorbed photon leads to one of these processes is hard to estimate and it is therefore a challenge to derive the photo-destruction timescales. It is nonetheless possible to understand, in a qualitative sense, how the photo-destruction timescale evolves from one PDR to another. If the frequency of photon absorption increases (i.e. the absorption timescale $\tau_{\mathrm{abs}}$ decreases) and the energy of these absorbed photons increases as well, the nano-grain has more chances to be photo-destroyed. Finally, in PDRs where the energy of the absorbed photons is almost identical, the ratio of the photo-destruction timescales can be approximated by the ratio of the absorption timescales.


Prior to their photo-destruction, nano-grains must have time to be photo-processed\footnote{The photo-processing refers to the destruction of C-H bonds (leading to the progressive aromatisation of the grain due to the creation of C=C bonds) and/or the destruction of C-C and C=C bonds.}. Thus, we need to compare the absorption timescale to the advection timescale, $\tau_{\mathrm{ad}}$, that is, the time that the incident UV light needs to heat up and dissociate the molecular gas at the cloud border. In practice, $\tau_{\mathrm{abs}}(a)$ must be smaller than $\tau_{\mathrm{ad}}$ to allow nano-grains to be photo-processed. The advection timescale is defined as $\tau_{\mathrm{ad}}=L/v_{\mathrm{DF}}$ where $L$ is the typical scale over which the radiation field penetrates the PDR which we assume corresponds to the position of the maximum emission in the NIR (i.e. where nano-grains are emitting). Following \cite{goldshmidt_time-dependent_1995}, the velocity of the dissociation front $v_{\mathrm{DF}}$ can be defined as:

\begin{equation}
    \label{eq:vDF}
    v_{\mathrm{DF}} = 71 \times \left(\frac{G_0}{10^{3}}\right) \left(\frac{n_{\mathrm{H}}}{10^{4}\,\mathrm{H\,cm^{-3}}}\right)^{-1} \mathrm{km\,s^{-1},}
\end{equation}

\noindent where $n_{\mathrm{H}}$ is the gas density at the position of the maximum NIR emission (i.e. at a distance, $L$). We summarise the velocities and the advection timescales found in Table\,\ref{tab:resume1}. The advection timescales fall within the range $\tau_{\mathrm{ad}}\sim 10^{3}-10^{4}$ yr.

\begin{table*}[h]
    \centering
    \begin{tabular}{lccccc}
        \hline
        \hline
        PDR & $G_0$ & $n_{\mathrm{H}}$ & $L$ & $v_{\mathrm{DF}}$ & $\tau_{\mathrm{ad}}$ \\
        & & [\,H\,cm$^{-3}$\,] & [\,pc\,] & [\,km\,s$^{-1}$\,] & [\,yr\,] \\
         \hline
          IC63      & $1100\pm 100$                  & $(2\pm 0.2)\times 10^{4}$ & $(8\pm 0.05)\times 10^{-3}$   & $0.39\pm 0.081$ &  $(2\pm 0.4)\times 10^{4}$ \\
          Horsehead & $100\pm 10$                    & $(1\pm 0.5)\times 10^{4}$ & $(7\pm 2.5)\times 10^{-3}$    & $0.07\pm 0.038$ &  $(5\pm 3)\times 10^{4}$   \\
          Orion Bar & $(2.6\pm 1.5)\times 10^{4}$ & $(4\pm 0.1)\times 10^{4}$ & $(1\pm 0.01)\times 10^{-2}$   & $4.6\pm 2.6$    &  $(1.8\pm 1)\times 10^{3}$ \\
        \hline
    \end{tabular}
    \caption{Summary of the different parameters required to determine the velocity of the dissociation front, $v_{\mathrm{DF}}$ (see Eq.\,\eqref{eq:vDF}), and the advection timescale, $\tau_{\mathrm{ad}}$. The details on the calculation of the advection timescale uncertainties can be found in Appendix\,\ref{app:uncertainties}.}
    \label{tab:resume1}
\end{table*}

The timescale $\tau_{\mathrm{abs}}(a)$ between two photons absorption by a dust grain with a size $a$ is defined as:

\begin{equation}
    \tau_{\mathrm{abs}}(a) = \int_{\nu_{\mathrm{min}}}^{\nu_{\mathrm{max}}} \pi a^{2} Q_{\mathrm{abs}}(a,\,\nu)\,\frac{I_{\nu}}{h\nu}\mathrm{d}\nu,
\end{equation}

\noindent where $Q_{\mathrm{abs}}(a,\,\nu)$ is the absorption efficiency at frequency $\nu$ for a dust size $a$ and $I_{\nu}$ is the specific intensity of the incident radiation field.

\begin{figure}[h]
\centering
\includegraphics[width=0.5\textwidth, trim={0 0cm 0cm 0cm},clip]{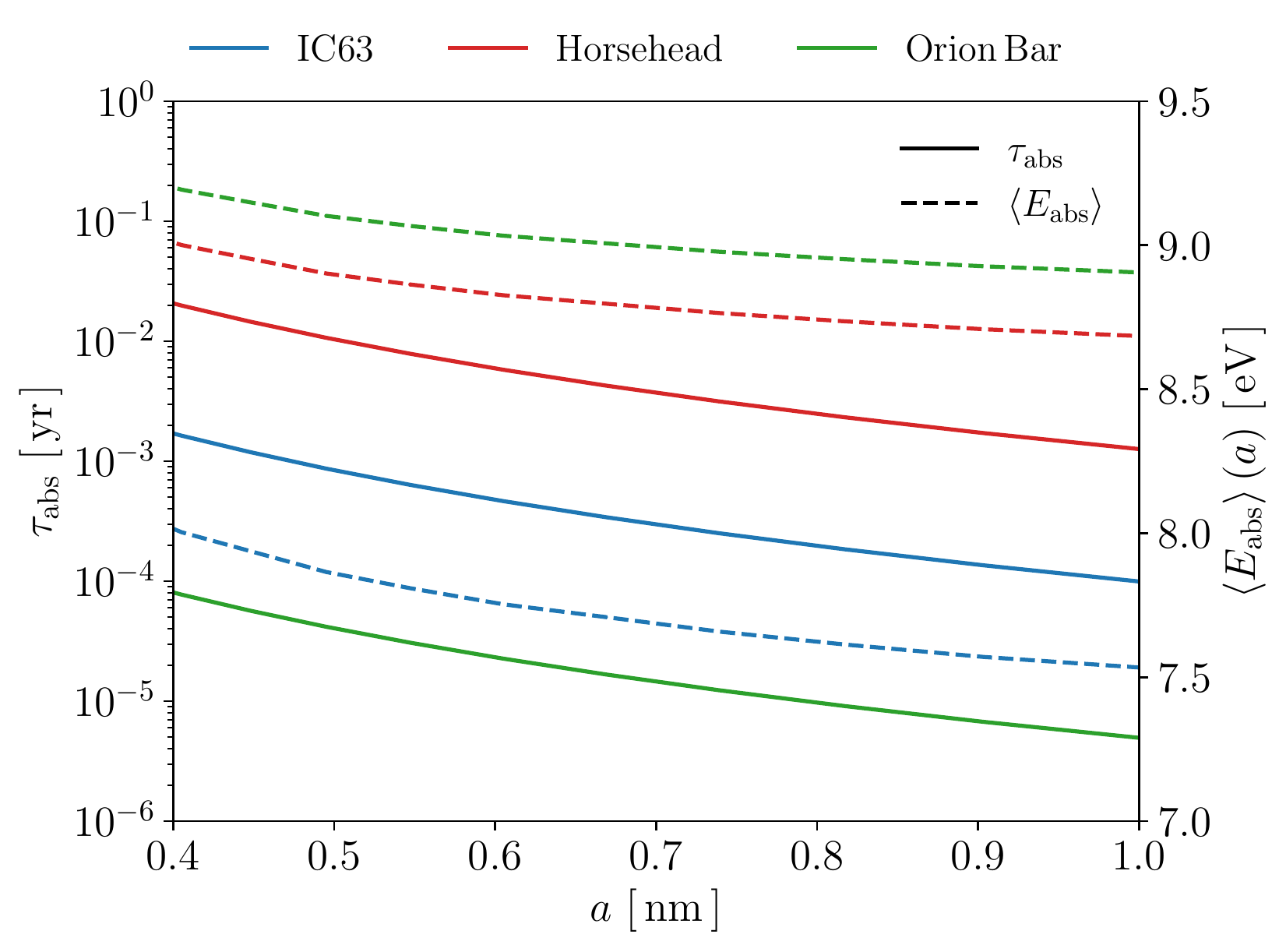}
    \caption{Absorption timescale (full line) and average energy of an absorbed photon (dashed-line) as a function of the dust size for the Horsehead (red), IC63 (blue), and the Orion Bar (green).}
    \label{fig:nathalie}
\end{figure}

Figure\,\ref{fig:nathalie} shows the absorption timescale as a function of the dust size for the Horsehead, IC63, and the Orion Bar. Regardless of the PDR and the dust size, these timescales are smaller than the advection timescales. We therefore assume that the nano-grains have time to be photo-processed in those three PDRs.

To assess whether dust grains are likely to be photo-destroyed in the Horsehead, IC63, and in the Orion Bar, we estimate the average energy of a photon which is absorbed by a dust grain, $\left<E_{\mathrm{abs}}\right>(a)=\left<h\nu_{\mathrm{abs}}\right>(a)$, as a function of the dust size and for the illumination conditions in those three PDRs (see Fig.\,\ref{fig:nathalie}). We observe that $\left<E_{\mathrm{abs, \,Orion}}\right> > \left<E_{\mathrm{abs,\,Horsehead}}\right> > \left<E_{\mathrm{abs,\,IC63}}\right>$. This is consistent with the star temperature being higher in the Orion Bar than in the Horsehead, which is itself higher than that in IC63 ($T_{\mathrm{star}}=38000,\,35000,\,25000$ K, respectively). This suggests that nano-grains are most efficiently photo-destroyed in the Orion Bar, then slightly less so in the Horsehead, and, finally, the least efficiently in IC63. Those energies are almost the same for the Orion Bar and the Horsehead, which suggests that the photo-destruction efficiency slightly changes between those two PDRs compared to IC63. We also observe that $\tau_{\mathrm{abs,\,Horsehead}}>\tau_{\mathrm{abs,\,IC63}}>\tau_{\mathrm{abs,\,Orion}}$ which is consistent with the intensity of the incident radiation field larger in the Orion Bar than in IC63, itself higher than in the Horsehead (\G~= 26000, 1100, 100, respectively). This suggests that nano-grains are more frequently photo-destroyed in the Orion Bar, then less so in IC63, and, finally, the least so in the Horsehead. The photo-destruction timescale in the Orion Bar is therefore the lowest among those three PDRs. Also, as the average energy of absorbed photons is almost identical in the Orion Bar and in the Horsehead, the ratio of the photo-destruction timescales in those two PDRs can be approximated by the ratio of the absorption timescales, which is $\tau_{\mathrm{abs,\,Horsehead}}/\tau_{\mathrm{abs,\,Orion}}\sim 100$. The photo-destruction timescale must therefore be about two orders of magnitude larger in the Horsehead than in the Orion Bar.  

Also, photo-destruction timescales are most likely proportional to the number of C-C and C=C bonds, itself proportional to the number of carbons in a dust grain. As this number increases with $a^{3}$, there is probably a threshold in carbon number above which dust grains are difficult to destroy. In addition, if the photo-dissociation of C-C bonds does not happen frequently, carbonaceous grains might have time to accrete other carbon atoms from the gas phase or going through internal re-structuration that will make them more difficult to destroy. Nevertheless, these mechanisms are the same for the three PDRs and therefore our comparative study of the direct photo-destruction timescales in these PDRs is still legitimate. 


\subsection{(Re-)Formation of nano-grains through the fragmentation of larger grains}\label{sect:sect:fragmentation}

Stars form in dense clouds, where dust grains are expected to be large aggregates \citep[e.g.][]{kohler_dust_2015} and, therefore, where nano-grains are no longer existing. The radiative feedback of these freshly formed stars on their parent dense environments creates photon-dominated regions. The presence of nano-grains in these regions that were previously only filled with dust aggregates indicates that those nano-grains are most likely formed by the fragmentation of aggregates. Different processes can lead to the fragmentation of large grains into smaller grains, such as collisions, absorption of energetic photons, and Coulomb fragmentation. As these two last processes are likely to be less efficient because absorbed photon energy is rapidly dissipated in a grain (on the order of the vibrational timescale $\sim 10^{-13}$ s) and because large grains will not be sufficiently charged in PDRs to undergo Coulomb fragmentation, we focus here on the fragmentation driven by grain collisions. 

In the framework of THEMIS, when the local gas density increases, large grains can form a second mantle either through accretion of C and H atoms, available in the gas phase or through coagulation of a-C nano-grains on the surfaces of larger grains. These grains are called core-mantle-mantle (CMM) grains. In denser regions, CMM grains coagulate together to form aggregates \citep{kohler_dust_2015} called aggregate-mantle-mantle (AMM) grains. The size distribution of AMM can be found in Fig.\,\textcolor{blue}{3} of Paper\,I. From now on, we use the AMM from THEMIS to model aggregates. 

\begin{figure*}[h]
\centering
    \includegraphics[width=0.9\textwidth, trim={0 0cm 0cm 0cm},clip]{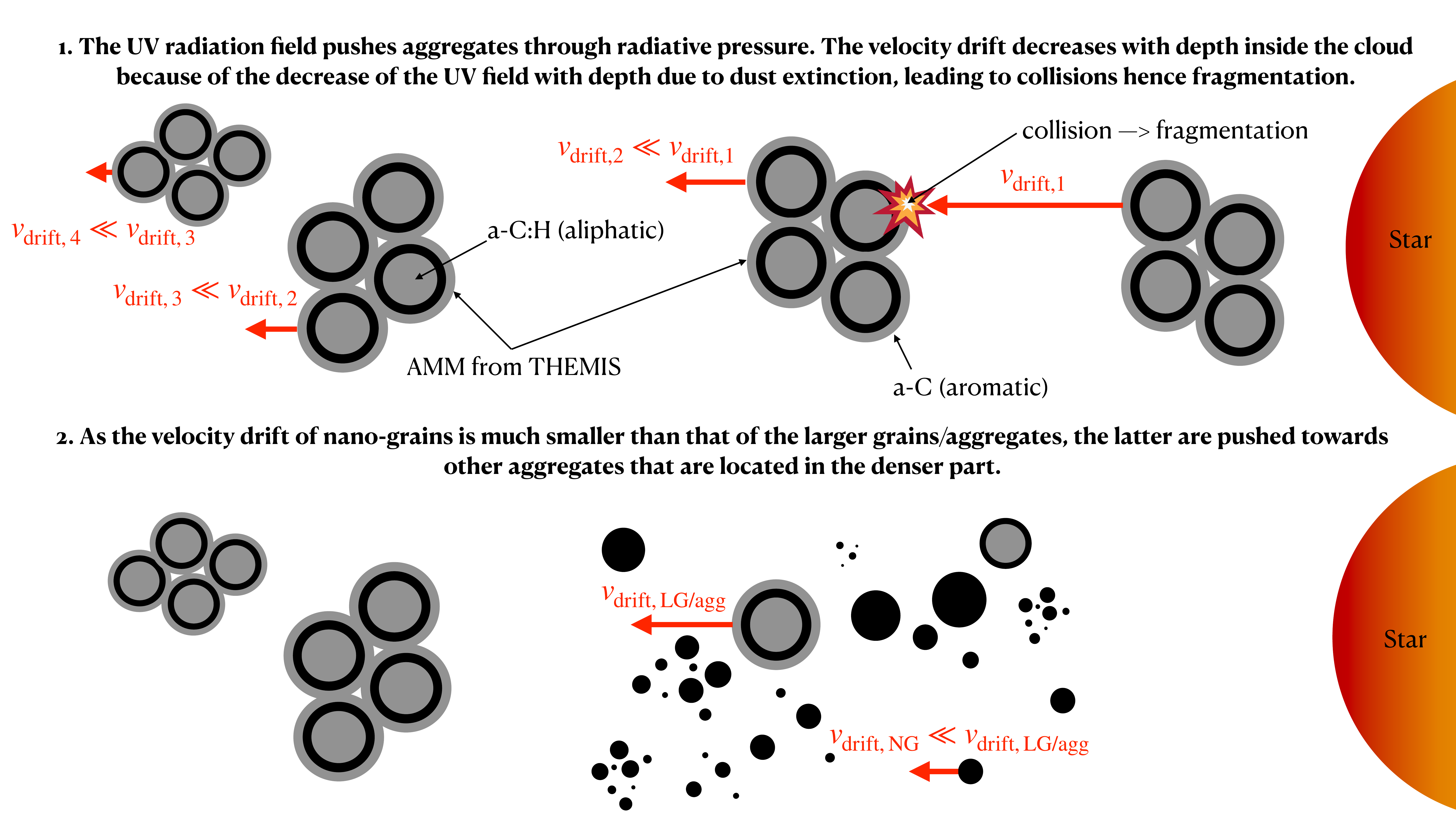}
    \includegraphics[width=0.9\textwidth, trim={0cm 37.5cm 0cm 0cm},clip]{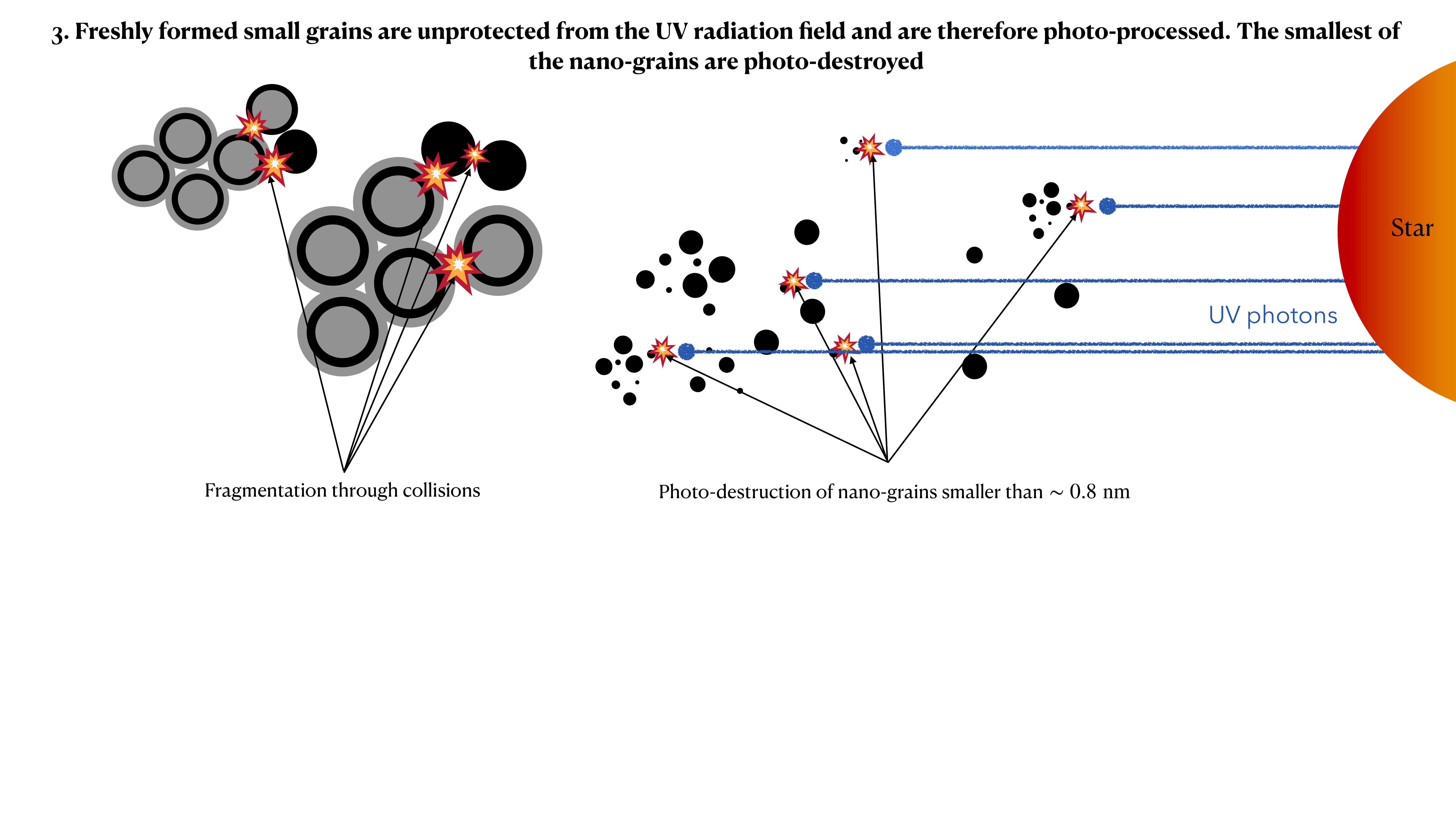}
\caption{Schematic view of the nano-grain formation through fragmentation driven by collisions of larger grains, followed by the nano-grain photo-destruction.}
    \label{fig:fragmentation_schema}
\end{figure*}

Using a classical one-dimensional approach, we estimate the drift velocities of aggregates driven by the radiative pressure caused by a blackbody with a temperature $T_{\star}$, $B_{\lambda}(T_{\mathrm{\star}})$, as well as the collision timescales associated. We consider an homogeneous medium filled with aggregates and gas, where those that are close to the star are pushed away. The gravitational force of the star on the dust grain can be non-negligible in some cases, for instance, in AGB environments \citep[e.g.][]{woitke_too_2006} and in circumstellar disks \citep[e.g.][]{vinkovic_radiation-pressure_2009, arnold_effect_2019}. We therefore consider the gravitational force in our model. In this depiction (see Fig.\,\ref{fig:fragmentation_schema}), an aggregate is subject to three forces:
\begin{enumerate}
    \item The radiative pressure force, $F_{\mathrm{pr}}(a)$, which is oriented along the star-aggregate axis;
    \item the force due to the drag caused by collisions with the gas, $F_{\mathrm{drag}}(a)$, oriented in the opposite direction to the radiative pressure force;
    \item the graviational force, $F_{\mathrm{grav}}(a)$, exerted by the illuminating star on the aggregate.
\end{enumerate}

The equation of motion of a dust aggregate with a mass, $m_{\mathrm{dust}}$, and a size, $a,$ is written as follows:

\begin{equation}
    \label{eq:1}
    m_{\mathrm{dust}} \frac{\mathrm{d}v_{\mathrm{dust}}(a)}{\mathrm{d}t} = F_{\mathrm{pr}}(a) - F_{\mathrm{drag}}(a) - F_{\mathrm{grav}}(a).
\end{equation}

\noindent The corresponding forces are defined as: 

\begin{equation}
    \label{eq:4}
    F_{\mathrm{drag}}(a) = \pi a^2 \,\rho_{\mathrm{gas}}\, v_{\mathrm{dust}}(a)\sqrt{v_{\mathrm{dust}}^2(a) + \frac{128\,k_{B} T_{\mathrm{gas}}}{9\pi\,\mu_{\mathrm{H}}}},
\end{equation}

\begin{equation}
    \label{eq:5}
    F_{\mathrm{grav}}(a) = \frac{4}{3}\pi a^3 \rho_{\mathrm{dust}} \times \left(\frac{G\,M_{\star}}{D^{2}}\right),
\end{equation}

\begin{equation}
    \label{eq:2}
    F_{\mathrm{pr}}(a) = \int_{\lambda} F_{\mathrm{pr}}(a,\,\lambda)\, \mathrm{d}\lambda,
\end{equation}

\noindent where 

\begin{equation}
    \label{eq:3}
    F_{\mathrm{pr}}(a,\,\lambda) = \pi a^2 Q_{\mathrm{pr}}(a,\,\lambda)\,\left(\frac{\pi R_{\star}^2}{D^2} \times \frac{ B_{\lambda}(T_{\mathrm{\star}})}{c} \right).
\end{equation}

In the equations above, the drag force $F_{\mathrm{drag}}$ is taken from \cite{jones_grain_1996} and the radiative pressure force, $F_{\mathrm{pr}}$, from \cite{abergel_evolution_2002}. In this force, the term in $v_{\mathrm{dust}}^{2}$ under the square root is associated with the intrinsic velocity of the dust grain while the second term corresponds to the gas particle velocity due to brownian motions. The different parameters are: $\rho_{\mathrm{gas}}=\mu_{\mathrm{H}}\,n_{\mathrm{H}}$ as the mean gas density, $\mu_{\mathrm{H}}=1.4\,m_{\mathrm{H}}$ as the mean atomic mass of the gas, $m_{\mathrm{H}}$ as the H atom mass, $T_{\mathrm{gas}}$ as the gas temperature, $\rho_{\mathrm{dust}}$ as the dust density, and $D$ is the distance from the aggregate to the star, while $n_{\mathrm{H}}$ is the gas density. The radiation pressure efficiency is defined as:

\begin{equation}
Q_{\mathrm{pr}}=Q_{\mathrm{abs}}+Q_{\mathrm{sca}}\times(1-g),
\end{equation}

\noindent where $Q_{\mathrm{abs}}$ ($Q_{\mathrm{sca}}$) is the dust absorption (scattering) efficiency and $g$ is the anisotropy factor of the scattering phase function.

In our case, the velocities of interest lead us to consider that the brownian motion of gas particles is, to first order, negligible compared to the dust motion hence Eq.\,\ref{eq:4} can be simplified to:

\begin{equation}
    \label{eq:6}
    F_{\mathrm{drag}}(a) = \pi a^2 \,\rho_{\mathrm{gas}}\, v_{\mathrm{dust}}^{2}(a).
\end{equation}

Using Eqs.\,\eqref{eq:5}, \eqref{eq:2}, and \eqref{eq:6} in Eq.\,\eqref{eq:1}, we obtain a first-order equation whose asymptotic solution corresponds to the drift velocity:

\begin{equation}
    v_{\mathrm{drift}}(a)=\frac{1}{D}\left[\frac{1}{1.4 m_{\mathrm{H}}\, n_{\mathrm{H}}} \left(\frac{\pi R_{\star}^{2}}{c}\left<Q_{\mathrm{pr}}B_{\lambda}\right>- \frac{4}{3}a\,\rho_{\mathrm{dust}}\,G M_{\star}\right) \right]^{1/2},    
\end{equation}

\noindent with

\begin{equation}
    \left<Q_{\mathrm{pr}}B_{\lambda}\right> = \int_{\lambda}Q_{\mathrm{pr}}(a,\,\lambda)\,B_{\lambda}(T_{\star})\,d\lambda.
\end{equation}

As we are interested in the regions of PDRs where nano-grains exist (i.e. where NIR and MIR dust emission reaches a maximum), we calculate the drift velocity for two extreme values of the gas density that are $n_{\mathrm{H}}=10^4$ and $10^5$ H\,cm$^{-3}$, which define the slice where fragmentation takes place. 

\begin{figure}[h]
\centering
\includegraphics[width=0.5\textwidth, trim={0 0cm 0cm 0cm},clip]{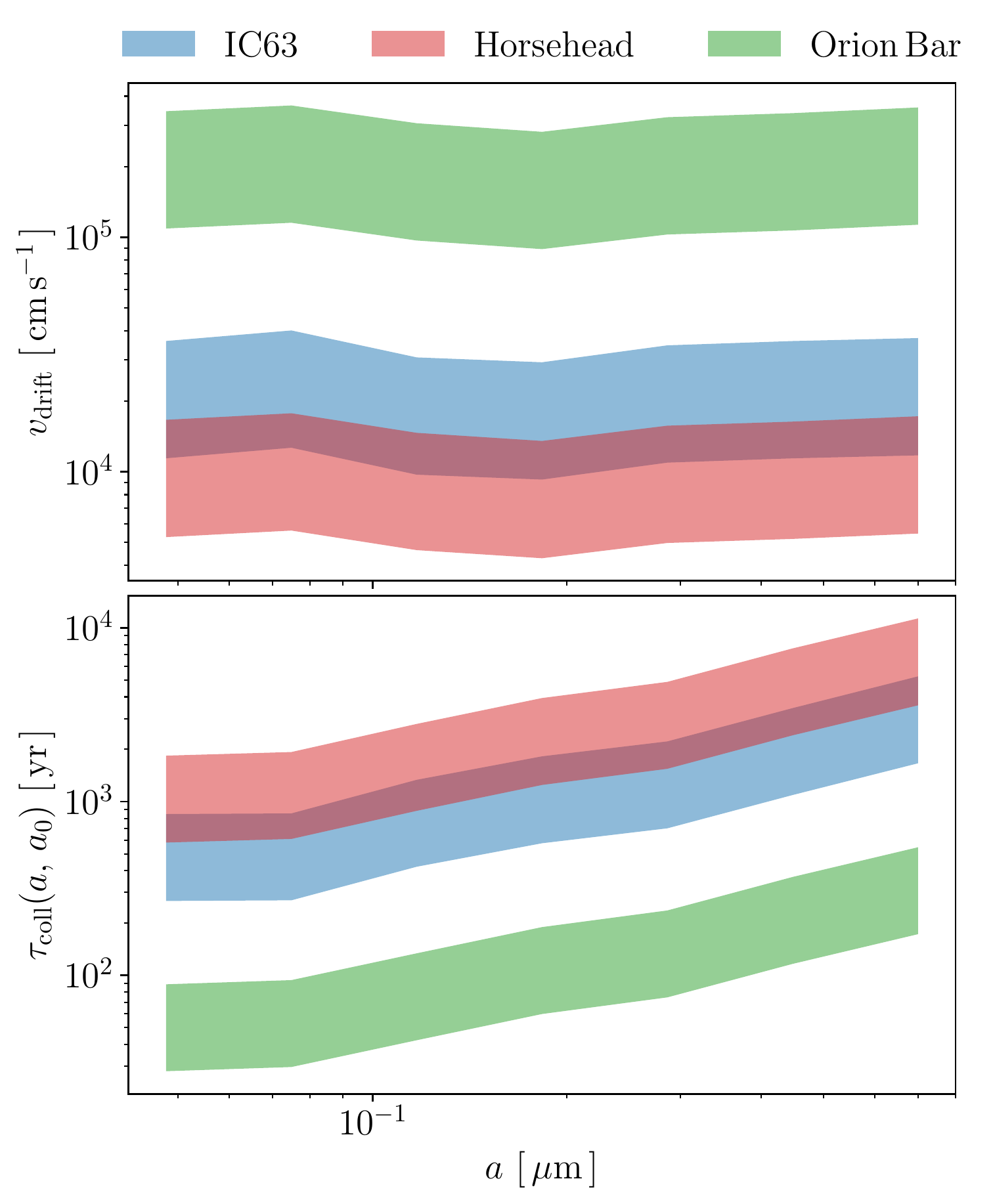}
    \caption{Drift velocities and collision timescales. Top: Drift velocities for IC63 (blue), Horsehead (red), and the Orion Bar (green) with \nh~varying from $10^{4}$ to $10^{5}$ H\,cm$^{-3}$. Bottom: Collision timescales between an AMM grain with size $a$ colliding with another AMM with size $a_0\sim 0.05$ \mum.}
    \label{fig:fragmentation}
\end{figure}

We show, in Fig.\,\ref{fig:fragmentation}, the drift velocities of AMM in IC63, the Horsehead, and the Orion Bar for \nh~varying from $10^{4}$ to $10^{5}$ H\,cm$^{-3}$. Regardless of the PDR, the drift velocity barely depends on the dust size (from 0.05 to 0.7 \mum). Collisions between two dust grains not only lead to fragmentation, they can also stick together and lead to grain growth. \cite{guttler_outcome_2010} found that when two aggregates with similar sizes collide with a velocity larger than $v_{\mathrm{frag}}=1$ m\,s$^{-1}$, they will  both be fragmented into smaller species according to a power-law size distribution $n(a)\propto a^{\gamma}$ with $\gamma$ that varies from -2.1 to -1.2. As regardless of the size and the PDR, the drift velocity is at least $10^4$ larger than $v_{\mathrm{frag}}$ (see Fig.\,\ref{fig:fragmentation}, top panel), we can assume that all collisions between AMM in IC63, the Horsehead, and the Orion Bar, lead to their fragmentation. It is also important to note that $\gamma$ varies from -2.1 to -1.2, which means that aggregates are most likely fragmented into many very small grains rather than a few small grains.  

From there, it is possible to estimate the fragmentation timescale from the collision timescale $\tau_{\mathrm{coll}}(a_1,\,a_2)$ (i.e. the collision time between two aggregates of sizes $a_1$ and $a_2$) defined as follows:

\begin{equation}
    \tau_{\mathrm{coll}}(a_1,\,a_2) = \left(n_{\mathrm{H}}\sqrt{n_{\mathrm{d}}(a_1) n_{\mathrm{d}}(a_2)}\,\pi (a_{1}+a_{2})^{2}v_{\mathrm{rel}}(a_1,\,a_2)\right)^{-1},  
\end{equation}

where $n_{\mathrm{d}}(a)$ is the relative abundance of AMM grains with size $a$ and $v_{\mathrm{rel}}(a_1,\,a_2)$ the relative velocity between two AMM. As the two AMM grains are pushed inwards the same direction and because the velocity of a grain located deeper in the PDR is much lower\footnote{Due to dust extinction.} than the grain closer to the star, the relative velocity between the two grains is: $v_{\mathrm{rel}}(a_1,\,a_2)\simeq v_{\mathrm{drift}}(a_1)$.

We show in Fig.\,\ref{fig:fragmentation} (bottom panel) the collision timescales between an AMM grain with size $a$ and an AMM grain with size $a_0\sim 0.05$ \mum, the most abundant AMM. As the collision timescales are lower than the advection timescales, the fragmentation of large grains is a viable scenario for the formation of nano-grains in PDRs.

\subsection{Dust evolution scenario}\label{sect:sect:dust_evolution}

The fragmentation of larger grains due to collisions caused by radiative pressure appears to be an efficient mechanism to form nano-grains. The efficiency of this mechanism mostly depends on \G~whereas barely depends on the temperature of the illuminating star. We therefore expect the nano-grain formation to be more efficient in IC63 than in the Horsehead, and then more efficient in the Orion Bar than in IC63. On the other hand, the nano-grain photo-destruction efficiency depends on both the star temperature and \G. Conversely to the Orion Bar and the Horsehead, the average energy of the incident photons in IC63 is not large enough to efficiently photo-destroy nano-grains. We explain this scenario in detail here. 

\subsubsection{Dependence of the nano-grain formation and destruction efficiencies with PDR depth}\label{sect:sect:spatial}

Because of dust extinction, the UV radiation field decreases with depth inside the PDR, which affects the radiative pressure and therefore the nano-grain formation. Also, as the nano-grain destruction depends on the amount of photons, the destruction of nano-grains is strongly affected by dust extinction. In order to easily compare one PDR to another, we explain the following in terms of $A_{\mathrm{V}}$, which additionally takes the density profile into account. 

\begin{figure}[h]
\centering
\includegraphics[width=0.5\textwidth, trim={0 0cm 0cm 0cm},clip]{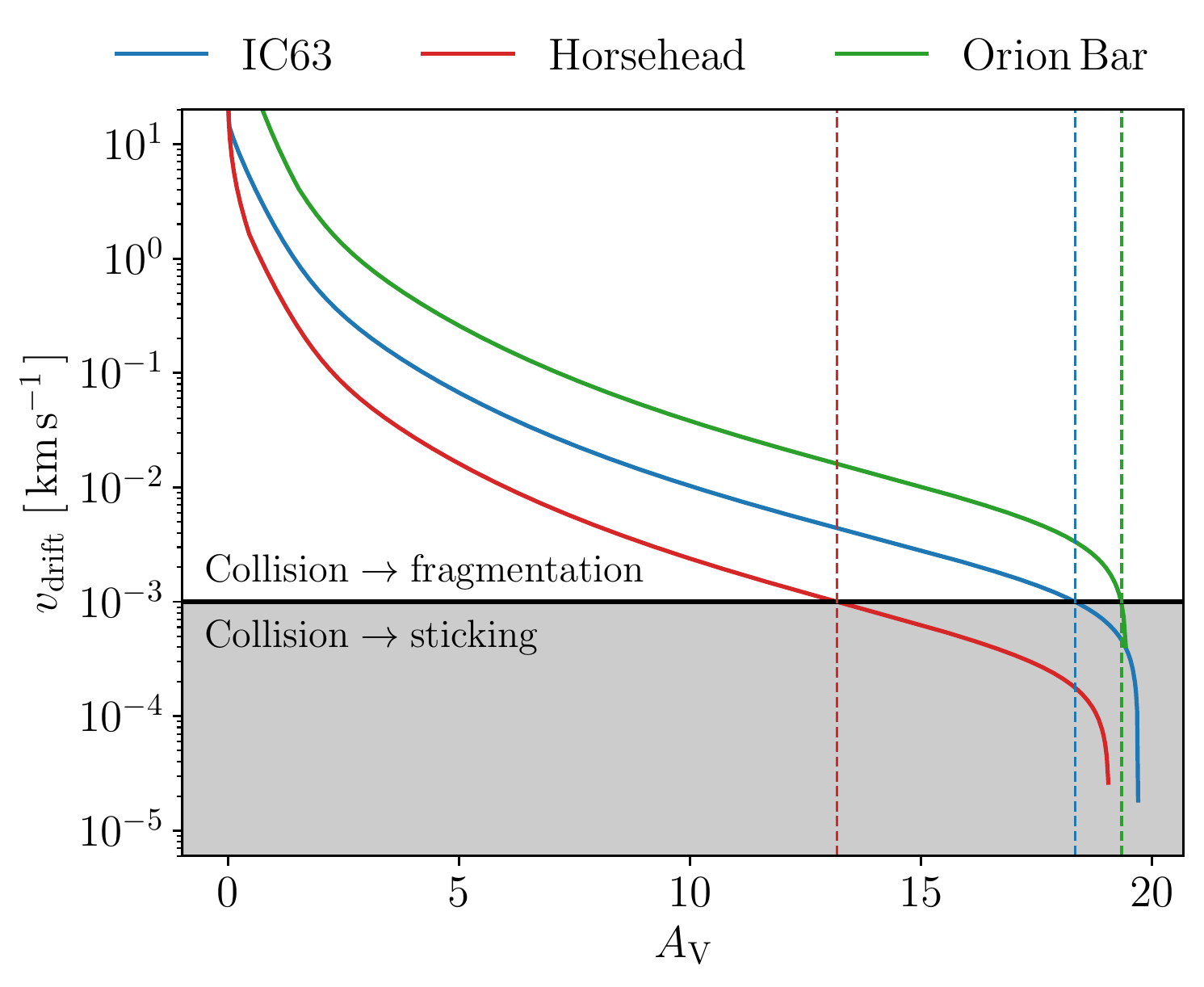}
    \caption{Drift velocity as a function of $A_{\mathrm{V}}$ for IC63 (blue line), Horsehead (Red line), and Orion Bar (green line). The horizontal black line corresponds to the velocity threshold above which collisions lead to fragmentation (see Sect.\,\ref{sect:sect:fragmentation}). The vertical dashed lines correspond to the limit in $A_{\mathrm{V}}$ above which the velocity drift is not high enough to generate fragmentation.}
    \label{fig:width1}
\end{figure}

We show, in Fig.\,\ref{fig:width1}, the drift velocity for the three PDRs as a function of $A_{\mathrm{V}}$. We see that regardless of the PDR, the velocity drift is greater than $v_{\mathrm{lim}}\sim 1$ m\,s$^{-1}$ (velocity above which collisions between dust grains lead to fragmentation) until $A_{\mathrm{V}}\sim 13.5$ ($A_{\mathrm{V,\,lim}}\sim 13.5, 18.5,\,\mathrm{and}\,19.5$ for the Horsehead, IC63, and the Orion Bar respectively). The net decrease in $v_{\mathrm{drift}}$ for large values of $A_{\mathrm{V}}$ is due to the fact that the radiative pressure force is lower than the gravitational force above these values of $A_{\mathrm{V}}$ hence the grain is attracted to the star (i.e. negative velocity drifts, not showed in this figure). This explains why we find aggregates in denser parts of PDRs and also why they are not all affected by fragmentation due to collisions. 

\begin{figure*}[h]
\centering
\includegraphics[width=\textwidth, trim={0 0cm 0cm 0cm},clip]{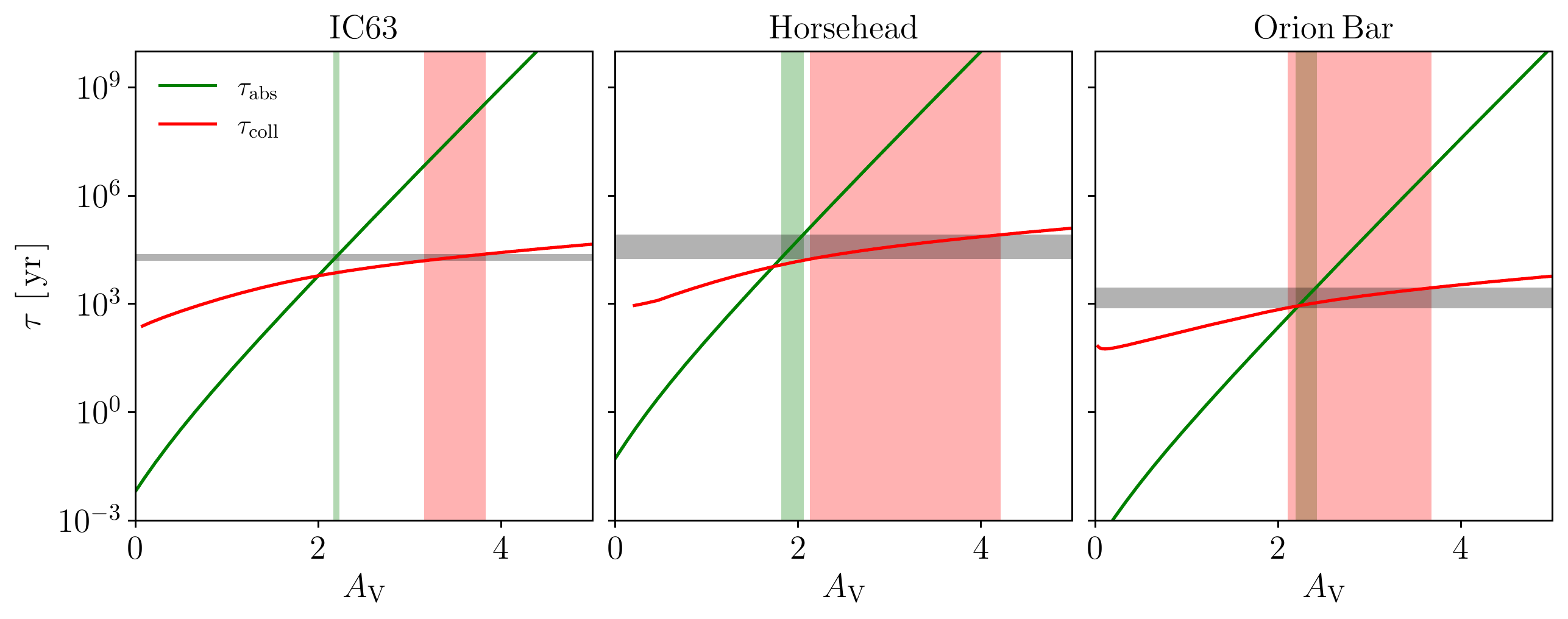}
    \caption{Absorption timescales are represented by the green lines for IC63 (left panel), Horsehead (middle panel), Orion Bar (right panel). Collision timescales between the most abundant aggregates ($a_0 \sim 0.05$ \mum) are shown in red lines. The horizontal grey stripes correspond to the advection timescales (see Table \ref{tab:resume1}). The vertical light green (red) stripes correspond to the region where $\tau_{\mathrm{abs}}\sim \tau_{\mathrm{ad}}$ ($\tau_{\mathrm{coll}}\sim \tau_{\mathrm{ad}}$). }
    \label{fig:width_subplots}
\end{figure*}

We show, in Fig.\,\ref{fig:width_subplots}, the collision timescales between the most abundant aggregates (AMM with $a_0\sim$ 0.05 \mum) for the three PDRs. If these timescales are larger than the advection timescales, collisions should not occur. This provides us with an upper limit on $A_V$ above which formation of nano-grains through the fragmentation is no longer efficient. These limits are $A_{\mathrm{V,\,lim}}\sim 3.1-3.9,\,2.1-4.2,\,\mathrm{and}\,2.1-3.7$ for IC63, Horsehead, and the Orion Bar, respectively. In these ranges, the drift velocity is larger than the velocity threshold, hence all collisions lead to fragmentation (see Fig.\,\ref{fig:width1}). Also, it means that above these values, aggregates must exist -- which is consistent with the existence of aggregates above $A_{\mathrm{V}}\geq 2-3$ \citep{ysard_variation_2013, ysard_mantle_2016}. Most importantly, the nano-grain maximum emission occurs for $A_{\mathrm{V}} < 0.5$ and because our scenario of nano-grain formation through fragmentation of aggregates is efficient below $A_{\mathrm{V}}\sim 2.1-3.9$ (depending on the PDR), this reinforces our scenario of nano-grain formation.  

We also show, in Fig.\,\ref{fig:width_subplots}, the absorption timescales. In order to be photo-destroyed, nano-grains must have time to absorb photons. In other words, $\tau_{\mathrm{abs}}$ must be lower than the advection timescale. Based on this, we obtain limits in $A_{\mathrm{V}}$ above which nano-grains are unlikely to be photo-destroyed. We note that this upper limit is therefore set at lower values of $A_\mathrm{V}$ as only a small fraction of the absorbed photons leads to the photo-destruction of nano-grains. Nevertheless, these limits are $A_{\mathrm{V,\,lim}}\sim 2.1-2.2,\,1.85-2.05,\,\mathrm{and}\,2.1-2.2$ for IC63, Horsehead, and the Orion Bar, respectively. As nano-grain maximum emission happens for $A_{\mathrm{V}} < 0.5$ and because nano-grains can be photo-processed below $A_{\mathrm{V}}\sim 1.85-2.2$ (depending on the PDR), they are most likely subject to photo-destruction below this threshold in $A_{\mathrm{V}}$. The mechanisms at the origin of the nano-grain formation and destruction are thus expected to be efficient where nano-grains emit in the irradiated outer part of PDRs.

\subsubsection{Evolution of the nano-grain abundance}\label{sect:sect:nano}

As nano-grains are efficiently formed in IC63 but not efficiently photo-destroyed, compared to the Horsehead, they are expected to be more abundant in IC63 than in the Horsehead. Regarding the Orion Bar, the discussion is slightly more challenging. Compared to the Horsehead, nano-grains are more efficiently photo-destroyed but also more efficiently formed through the fragmentation of larger grains. To first order, we assume that the efficiency of the nano-grain formation is proportional to the collision timescale and that the efficiency of the nano-grain destruction is proportional to the absorption timescale. We note that regardless of the dust size, the collision timescale is a factor $\sim 10$ larger in the Horsehead than in the Orion Bar (see Fig.\,\ref{fig:fragmentation}, bottom panel), whereas the absorption timescale is a factor $\sim 100$ greater in the Horsehead than in the Orion Bar (see Fig.\,\ref{fig:nathalie}). This means that the ratio between the destruction efficiency and the formation efficiency is $\sim 10$ times greater in the Orion Bar than in the Horsehead. This supports our results as we find that the nano-grain abundance is lower by a factor of at least 10 in the Orion Bar compared to the Horsehead, which suggests that the ratio between the mechanism that leads to the nano-grain destruction with the one leading to their formation shown to be greater in the Orion Bar than in the Horsehead. 

\subsubsection{Evolution of the nano-grain minimum size and power-law size-distribution exponent}\label{sect:sect:amin}

In the diffuse ISM, the nano-grain minimum size is around 0.4 nm ,\citep{jones_evolution_2013} whereas it increases in PDRs to $\sim 0.7$ nm in IC63, $\sim 0.8$ nm in the Horsehead (see Paper\,I), and up to a maximum of $\sim 0.8$ nm in the Orion Bar. This increase of 0.1 nm in the nano-grain minimum size from IC63 to the Orion Bar and Horsehead implies an noticeable increase of 1.5 in the carbon atom number, which is expected as the average energy of absorbed photons in both the Orion Bar and Horsehead is larger than in IC63. Also, as the nano-grain minimum size varies little from one PDR to another, the photo-destruction timescale must drastically decrease above a given size limit around 0.7-0.8 nm. \cite{allain_photodestruction_1996} showed that despite the fact that the photo-destruction timescale barely varies from one small PAH to another (i.e. $N_{\mathrm{C}}<30$), it can vary broadly from a small PAH to a larger one ($N_{\mathrm{C}}\sim 50$). Indeed, photo-destruction timescales are expected to be about $10^{2}$ to $10^{3}$ years for PAHs going from benzene (C$_6$H$_6$) to ovalene (C$_{32}$H$_{14}$) while it can reach $\sim 10^{10}$ years for a PAH with 50 carbon atoms. As the photo-destruction of nano-grains depends on the number of C-C/C=C bonds and because this number evolves with $a^{3}$, one can understand that the photo-destruction timescales evolves quickly with $N_{\mathrm{C}}$ ($N_{\mathrm{C}}\propto a^{3}$).

In the diffuse ISM, the exponent of the THEMIS nano-grain power-law size distribution is $\alpha=-5$. In the Horsehead, we find a lower value $\alpha\sim -6$ and $\alpha\leq -6$ in the Orion Bar. This indicates that the fragmentation of large grains into nano-grains favours the formation of many small nano-grains instead of few larger nano-grains. The fact that we find the same value of $\alpha$ in IC63 as in the diffuse ISM supports the scenario where the formation efficiency of nano-grains through the fragmentation is larger than the photo-destruction efficiency. Indeed, this suggests that the nano-grains population had nearly enough time to completely reform in IC63, unlike the Horsehead and Orion Bar.

\section{Conclusion}\label{sect:conclusion}

We use \spitzer~and \herschel~data to map dust emission in two PDRs: IC63 and the Orion Bar. We modelled dust emission across those two PDRs using the THEMIS dust model together with the radiative transfer code SOC. We show that it is possible to partially constrain the gas density profile only based on the width of the dust emission profile. Using these density profiles, we show that dust similar to that of the diffuse ISM cannot explain the observations and, thus, the dust size distribution has to be modified. 

The nano-grain dust-to-gas mass ratio is roughly half that of IC63 and almost 100 times lower in the Orion Bar than in the diffuse ISM. The nano-grain minimum size is about 0.7 nm in IC63 and no more than 0.8 nm in the Orion Bar. The slope of the size distribution in IC63 is almost the same as in the diffuse ISM whereas it is steeper in the Orion Bar. This suggests that the mechanism at the origin of the nano-grain destruction is more efficient than the one of nano-grain formation in the Orion Bar compared to IC63. 

To this end, we estimate formation timescales assuming that nano-grains are mainly formed through collisions between larger grains driven by radiative pressure from the star. We also estimate destruction timescales assuming that nano-grains are destroyed by energetic photons. Based on this timescale analysis, we find that the nano-grain destruction-to-formation ratio increases from IC63 to the Orion Bar, through the Horsehead, which explains the decrease in the nano-grain abundance from IC63 to the Orion Bar. As the photo-destruction efficiency is quite low in IC63 compared to the Orion Bar and Horsehead, whereas the formation of nano-grains in IC63 is efficient, this explains why the presence of dust in IC63 is more similar to what we would find in the diffuse ISM -- as compared to what we would find in the Orion Bar and the Horsehead. The fact that the nano-grain minimum size scarcely varies from one PDR to another, but is still twice as large as than in the diffuse ISM suggests that there is a critical size above which dust grains are resilient to photo-destruction, regardless of irradiation. Based on our constraints on the nano-grain minimum size, we find that this critical size is around 0.7-0.8 nm.

\begin{acknowledgements} We thank the anonymous referee for very helpful suggestions and comments. This work was supported by the Programme National “Physique et
Chimie du Milieu Interstellaire” (PCMI) of CNRS/INSU with INC/INP co-funded
by CEA and CNES. We would like to thanks Olivier Berné for stimulating discussions on dust photo-dissociation. T.S. acknowledges support from the Knut and Alice Wallenberg Foundation (grant no. 2020.0081).
\end{acknowledgements}

\bibliographystyle{aa} 
\bibliography{aa}

\begin{appendix}
\section{IC63 and the Horsehead as seen with \spitzer~and \herschel}

\begin{figure*}[h]
\centering
        \includegraphics[width=0.33\textwidth, trim={0 0cm 0cm 0cm},clip]{IC63_0_new.pdf}\hfill
        \includegraphics[width=0.33\textwidth, trim={0 0cm 0cm 0cm},clip]{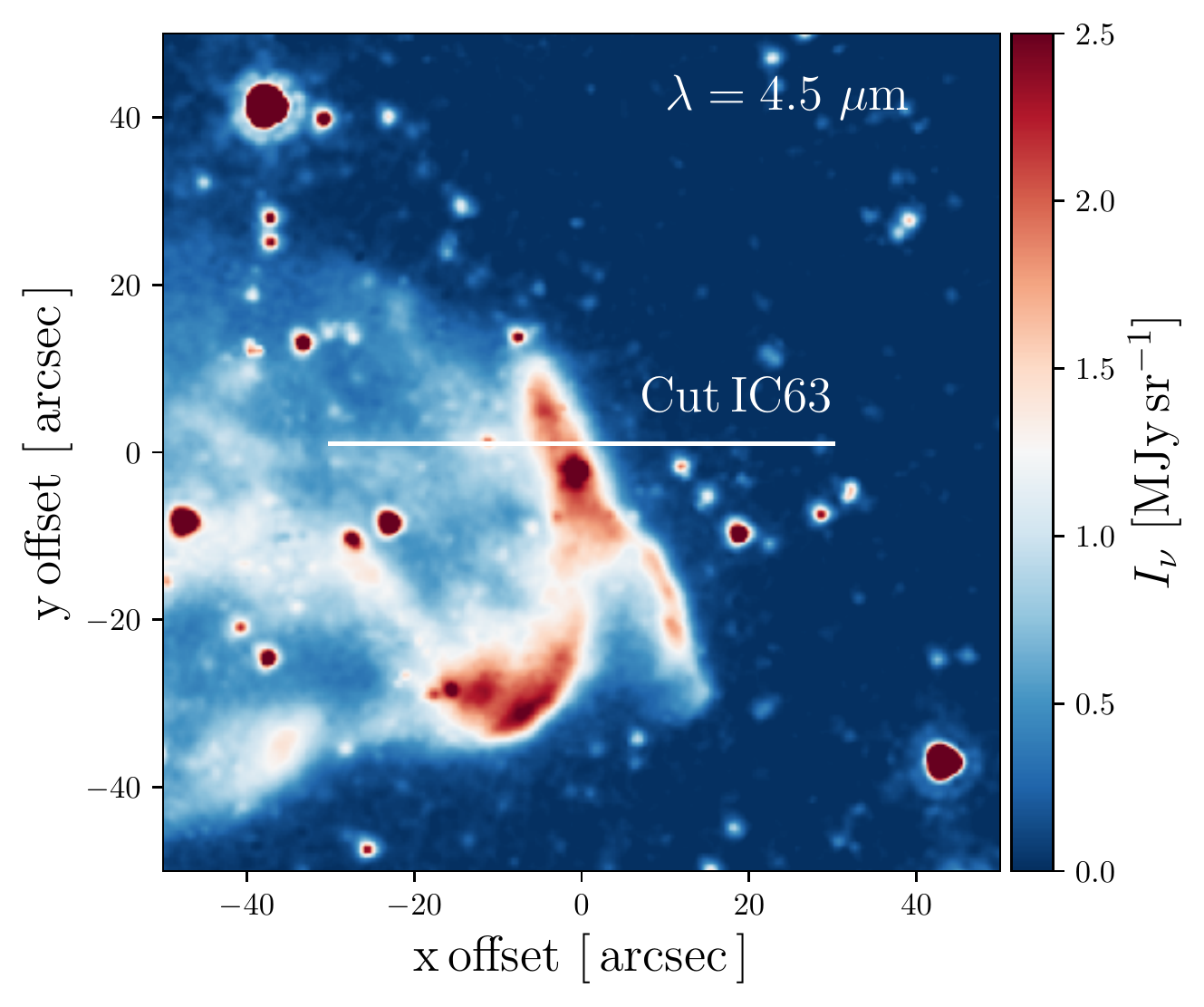}\hfill
        \includegraphics[width=0.33\textwidth, trim={0 0cm 0cm 0cm},clip]{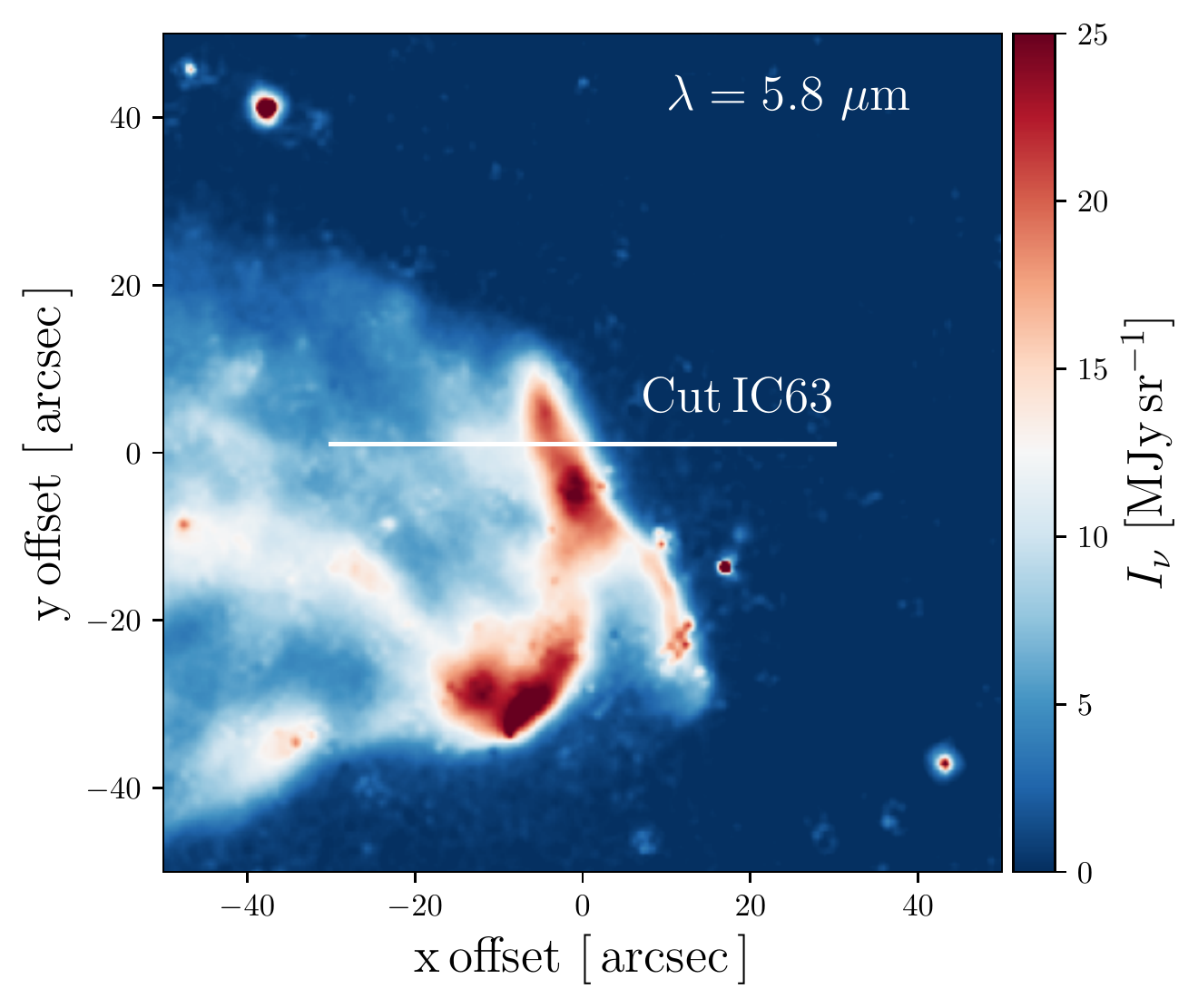}\\
        \includegraphics[width=0.33\textwidth, trim={0 0cm 0cm 0cm},clip]{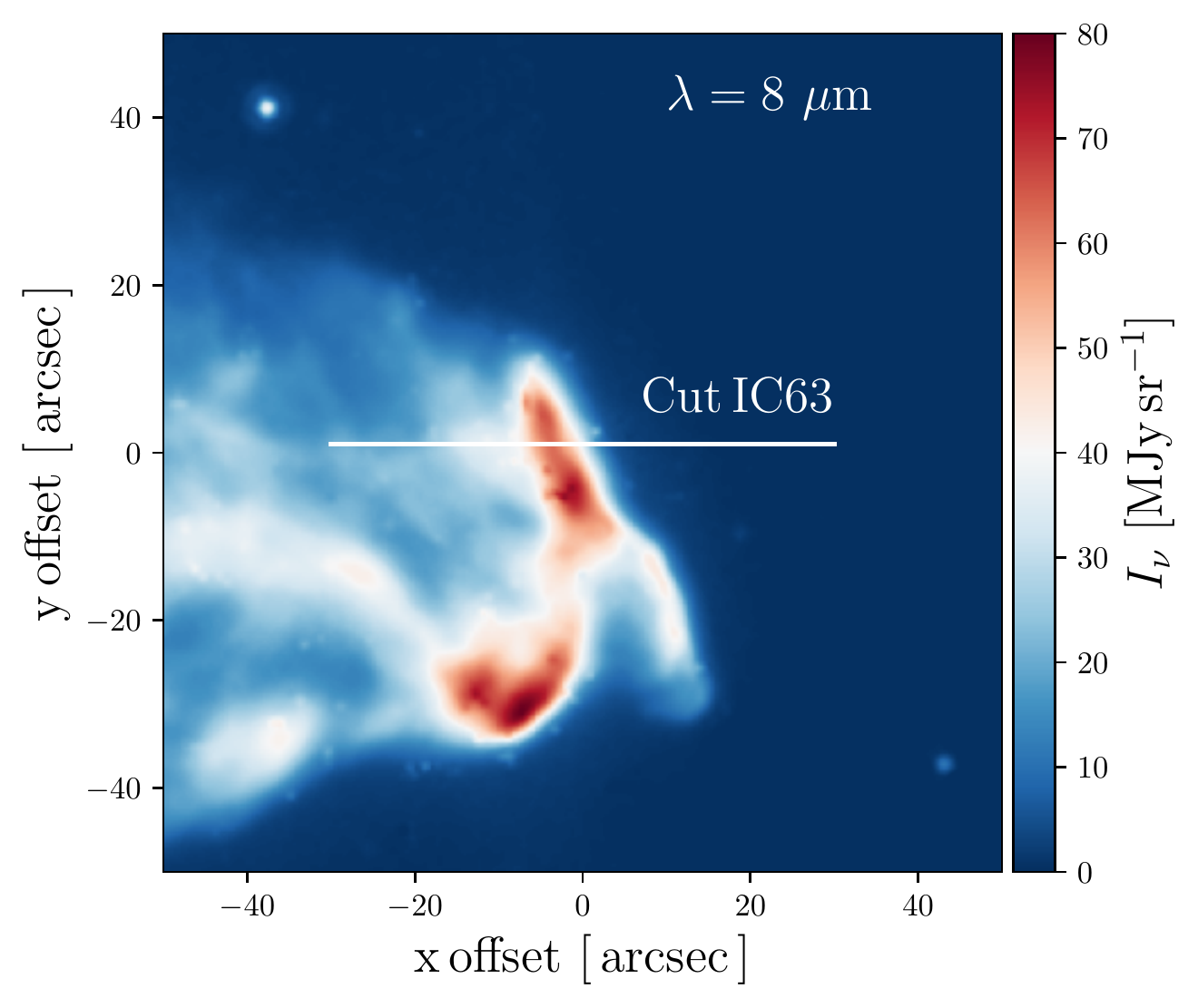}\hfill
        \includegraphics[width=0.33\textwidth, trim={0 0cm 0cm 0cm},clip]{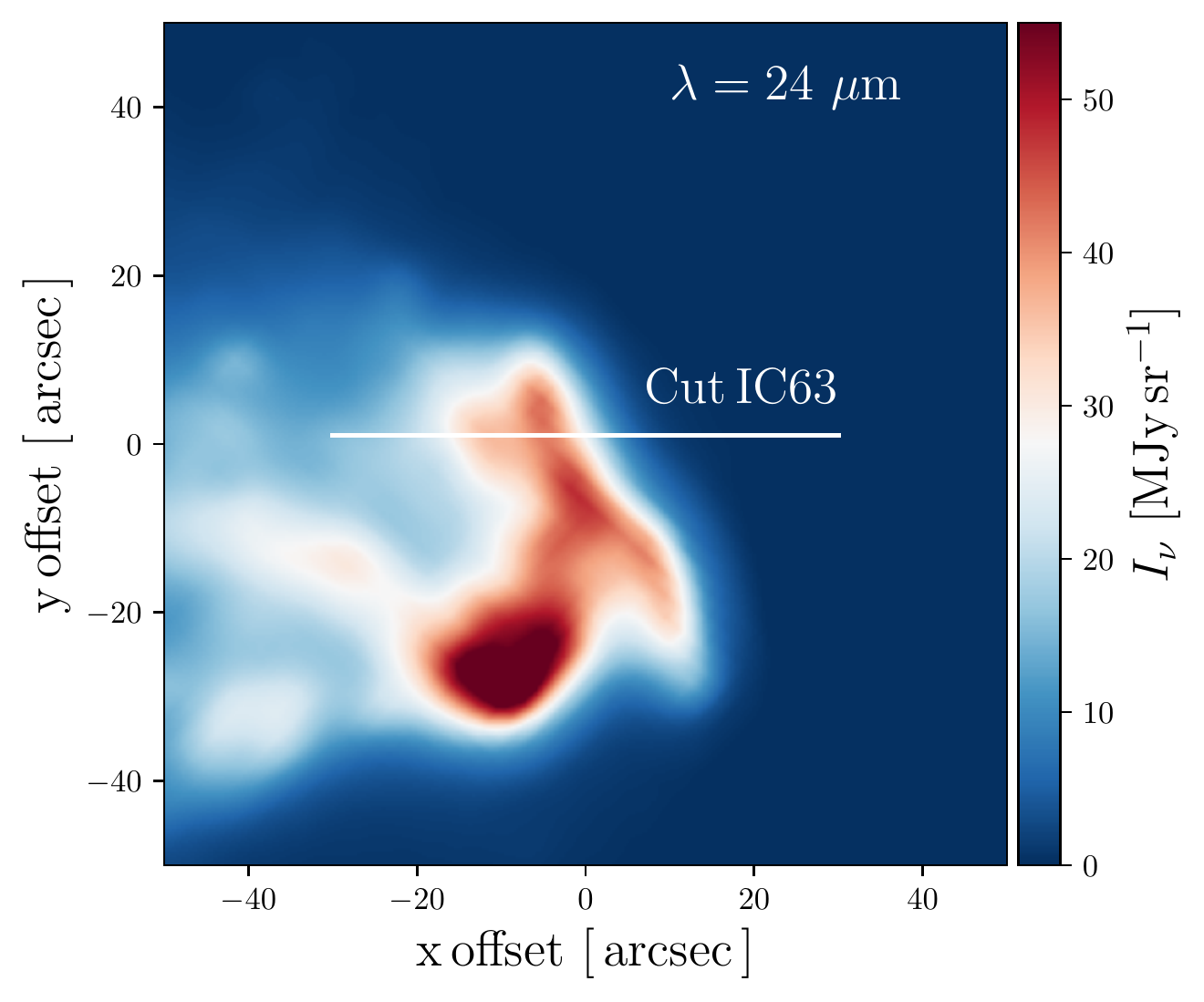}\hfill
        \includegraphics[width=0.33\textwidth, trim={0 0cm 0cm 0cm},clip]{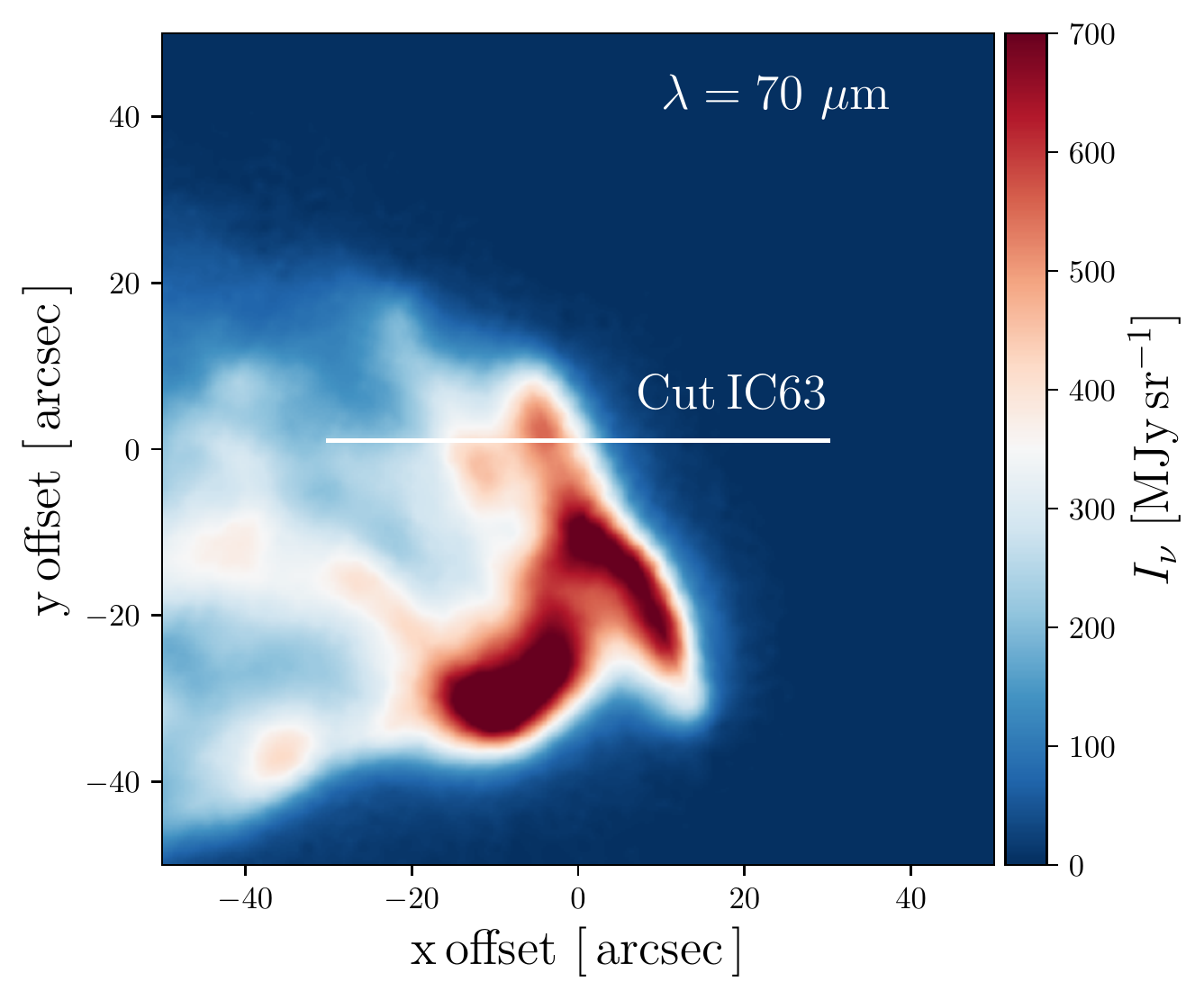}\\
        \includegraphics[width=0.33\textwidth, trim={0 0cm 0cm 0cm},clip]{orion_0.pdf}\hfill
        \includegraphics[width=0.33\textwidth, trim={0 0cm 0cm 0cm},clip]{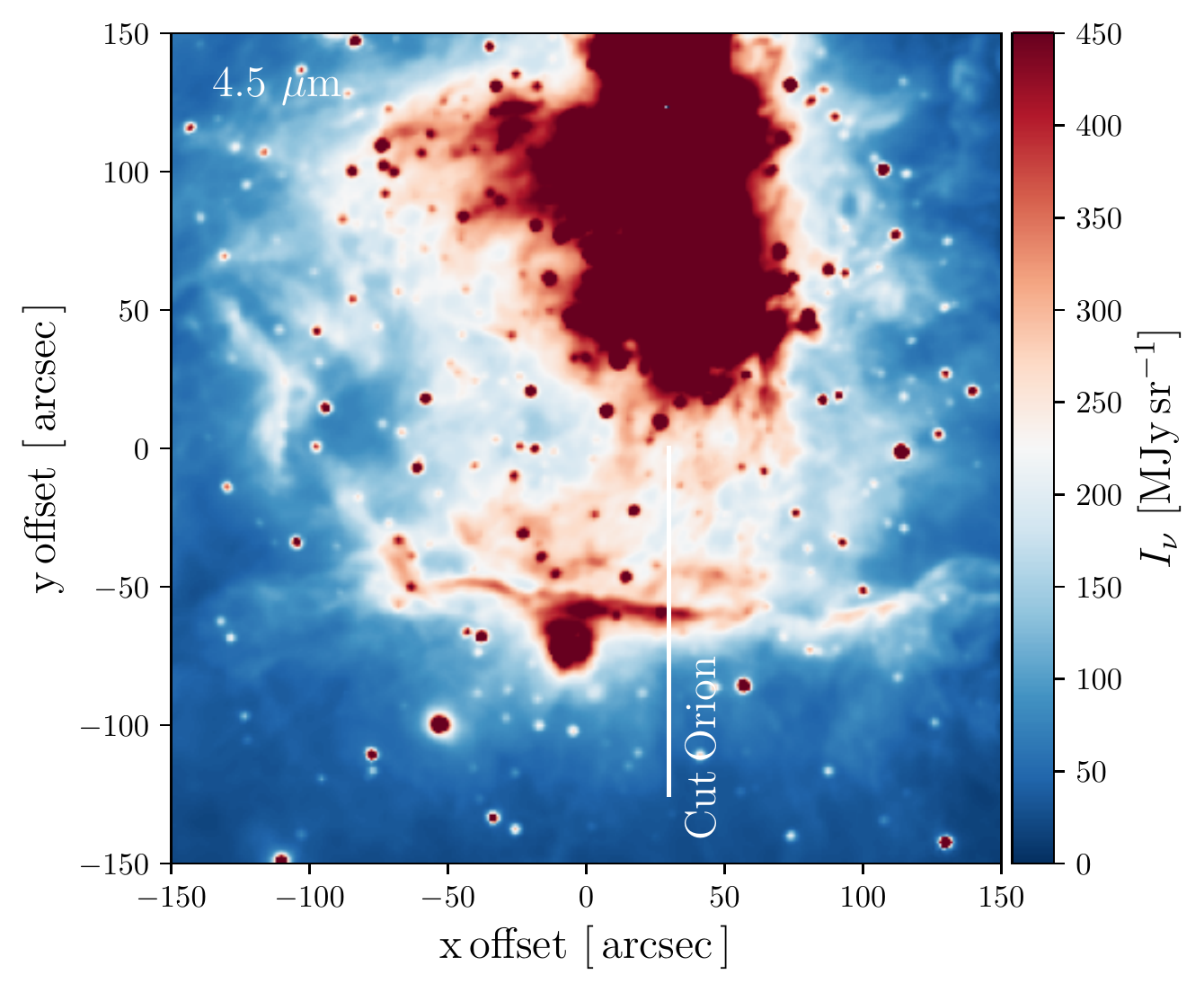}\hfill
        \includegraphics[width=0.33\textwidth, trim={0 0cm 0cm 0cm},clip]{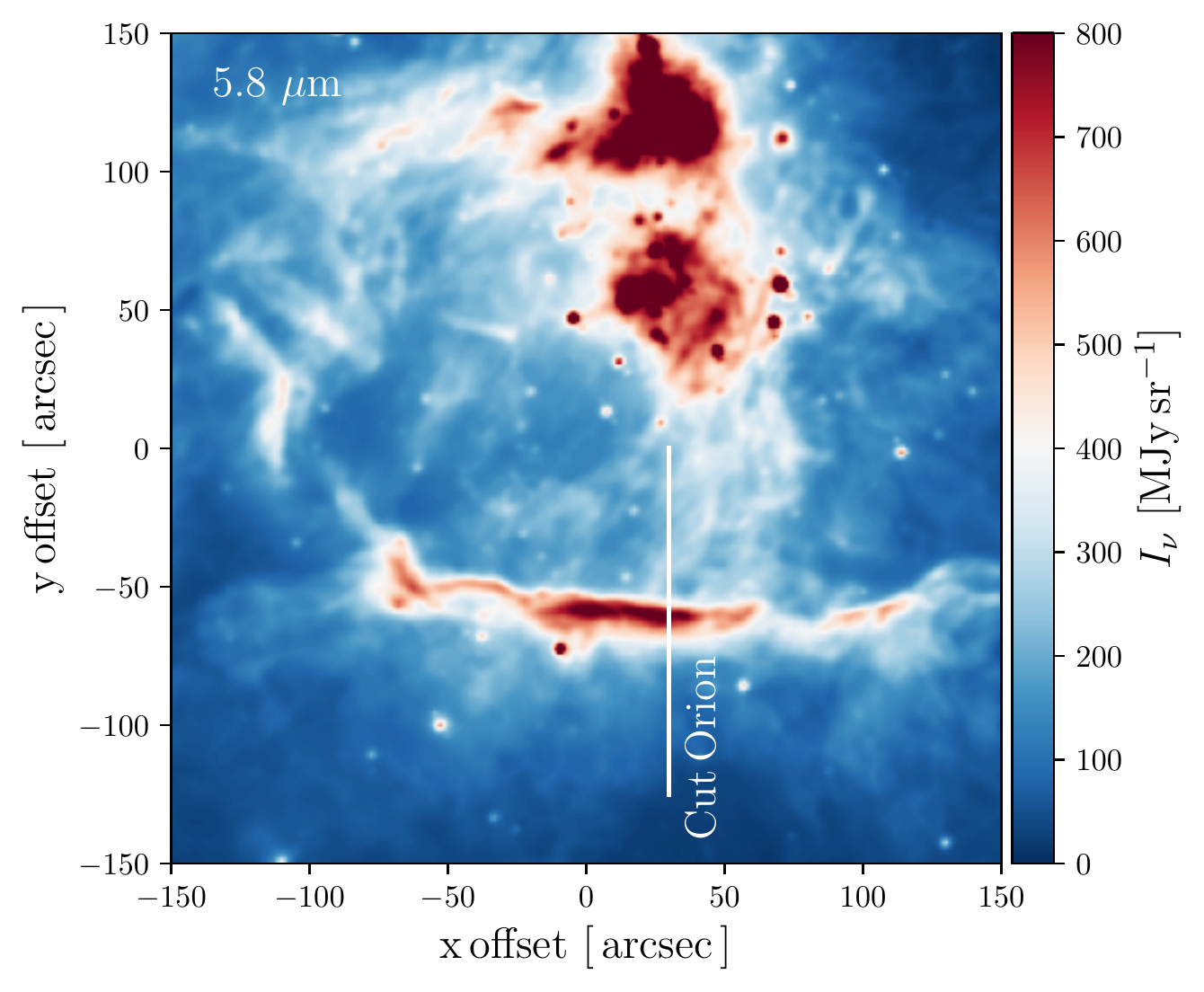}\\
        \includegraphics[width=0.33\textwidth, trim={0 0cm 0cm 0cm},clip]{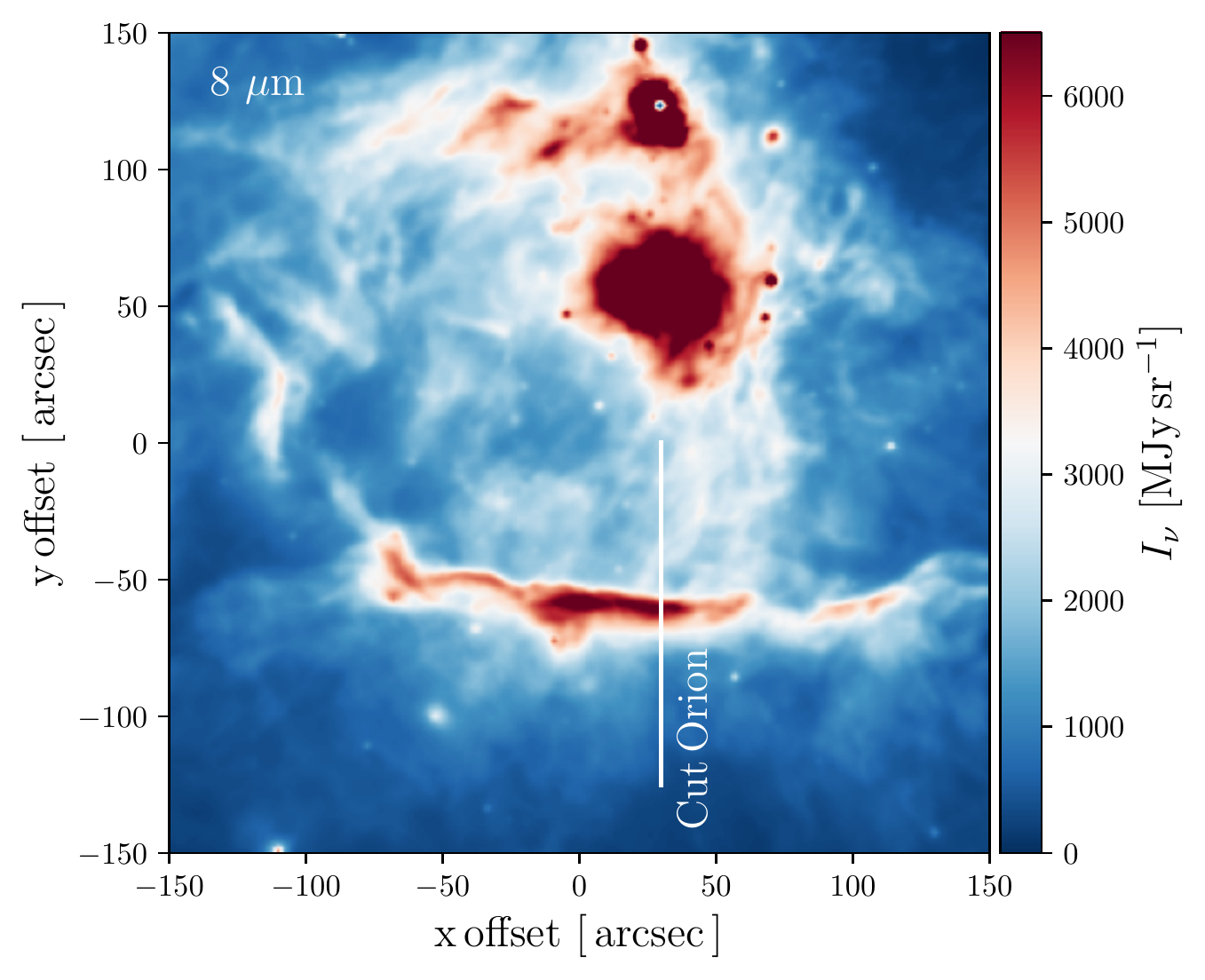}\hfill
        \includegraphics[width=0.33\textwidth, trim={0 0cm 0cm 0cm},clip]{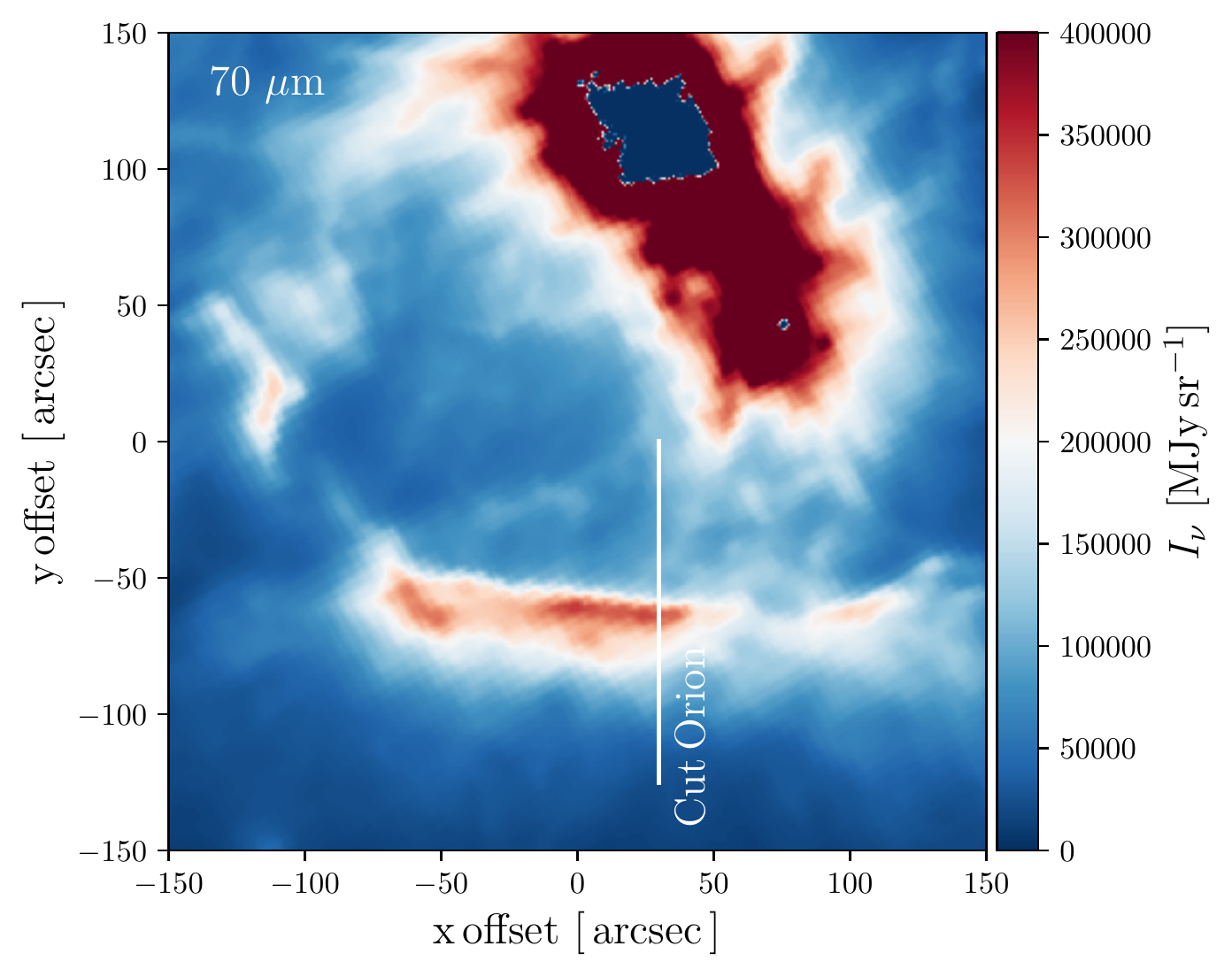}\hfill
    \caption{IC63 and the Orion Bar as seen with Spitzer and Herschel. Top six figures: IC63 seen in six photometric bands (3.6, 4.5, 5.8, 8, 24, and 70 \mum). Bottom five figures: the Orion Bar seen in five photometric bands (3.6, 4.5, 5.8, 8, and 70 \mum). The white solid
    lines correspond to the cuts used in our study.}
    \label{fig:IC63_orion_obs_all}
\end{figure*}

\section{Influence of variations in dust properties on the width of the emission profiles}
\label{appendix:width_emission}

\begin{figure}[h]
\centering
        \includegraphics[width=0.5\textwidth, trim={0 1.5cm 0cm 0cm},clip]{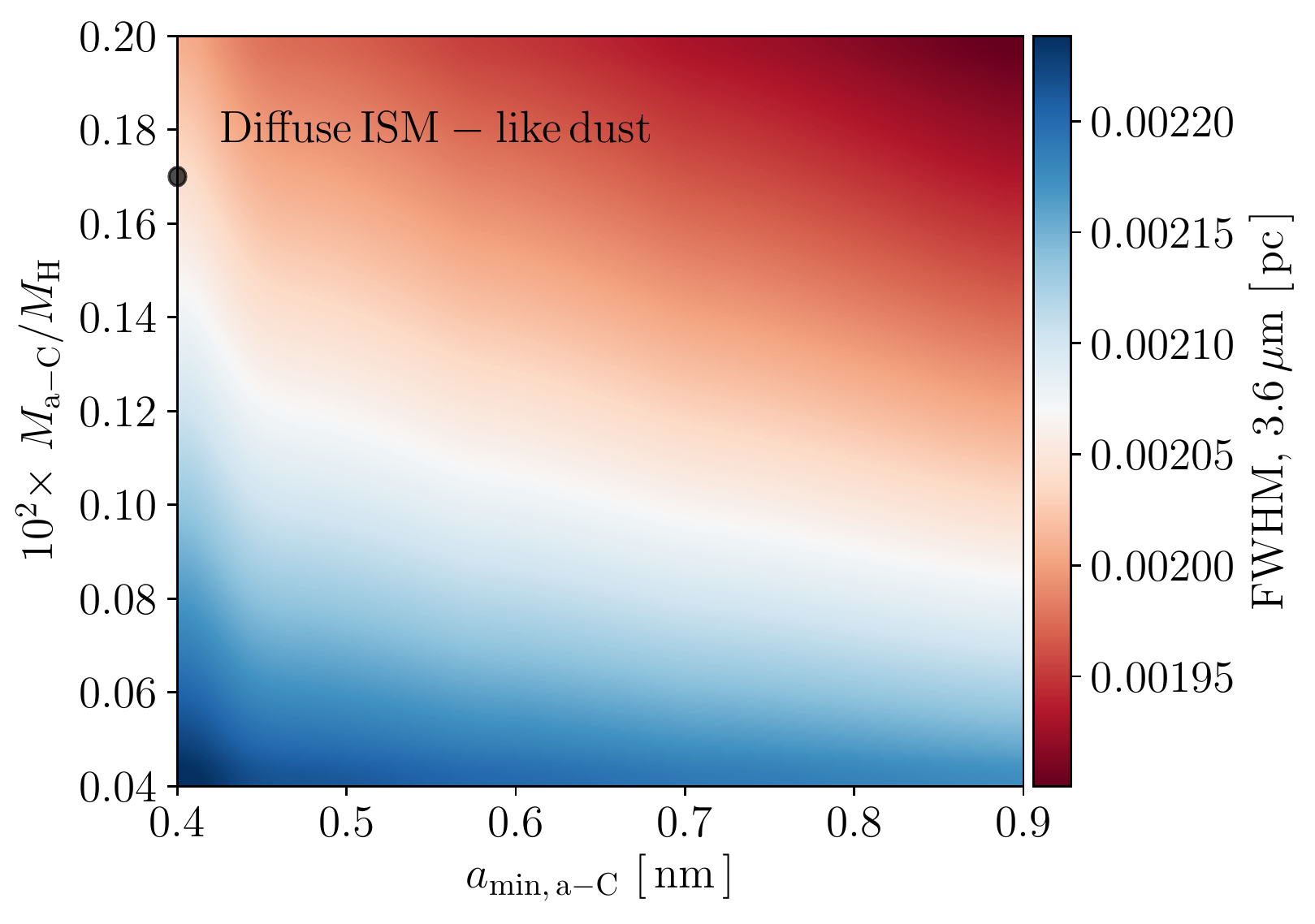}
        \includegraphics[width=0.5\textwidth, trim={0 0cm 0cm 0cm},clip]{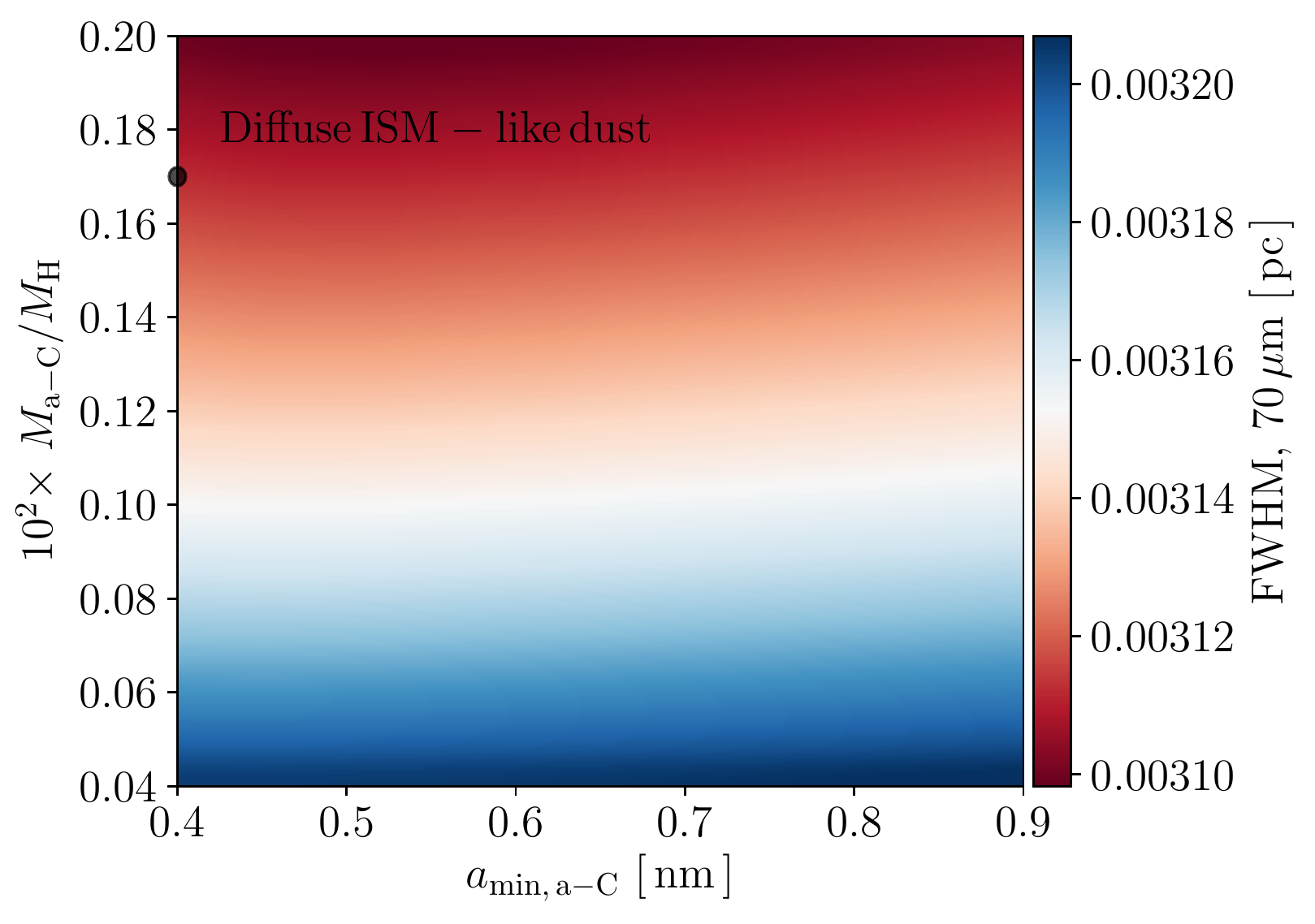}
\caption{FWMH in the 2D space (\abvsg, \aminvsg). Top: Full width at half maximum (FWHM) of the dust emission profile at 3.6 \mum~in the 2D space (\abvsg, \aminvsg). \textit{Bottom}: Same at 70 \mum.}
    \label{fig:amin_ab_filter05}
\end{figure}

Figure\,\ref{fig:amin_ab_filter05} shows the FWHM of the dust emission profile at 3.6 \mum~and 70 \mum~in the 2D space defined by \abvsg~and \aminvsg. 

\begin{table}[h]
    \centering
    \begin{tabular}{lccc}
        \hline
        \hline
        Band & std & min & max \\
         \hline
          3.6 \mum & $7.76\times 10^{-5}$ pc & $1.90\times 10^{-3}$ pc & $2.24\times 10^{-3}$ pc \\
          4.5 \mum & $7.32\times 10^{-5}$ pc & $1.94\times 10^{-3}$ pc & $2.22\times 10^{-3}$ pc \\
          5.8 \mum & $6.74\times 10^{-5}$ pc & $2.11\times 10^{-3}$ pc & $2.38\times 10^{-3}$ pc \\
          8 \mum & $6.34\times 10^{-5}$ pc & $2.25\times 10^{-3}$ pc & $2.49\times 10^{-3}$ pc \\
          24 \mum & $2.29\times 10^{-5}$ pc & $3.55\times 10^{-3}$ pc & $3.64\times 10^{-3}$ pc \\
          70 \mum & $3.08\times 10^{-5}$ pc & $3.10\times 10^{-3}$ pc & $3.2\times 10^{-3}$ pc \\
        \hline
    \end{tabular}
    \caption{Standard deviations, minima, and maxima of the FWHM - of the dust modelled emission - distribution in 2D space (\abvsg, \aminvsg) for the six following bands: 3.6, 4.5, 5.8, 8, 24, 70 \mum. The FWHM in 2D space (\abvsg, \aminvsg) at 3.6 and 70 \mum~are shown in Fig.\,\ref{fig:amin_ab_filter05}.}
    \label{tab:std_min_max}
\end{table}

\section{\chiTOT~minimisation in the 3D space (\abvsg, \aminvsg, and $\alpha$), IC63}

\begin{figure*}[h]
\centering
        \includegraphics[width=0.95\textwidth, trim={0 0cm 0cm 0cm},clip]{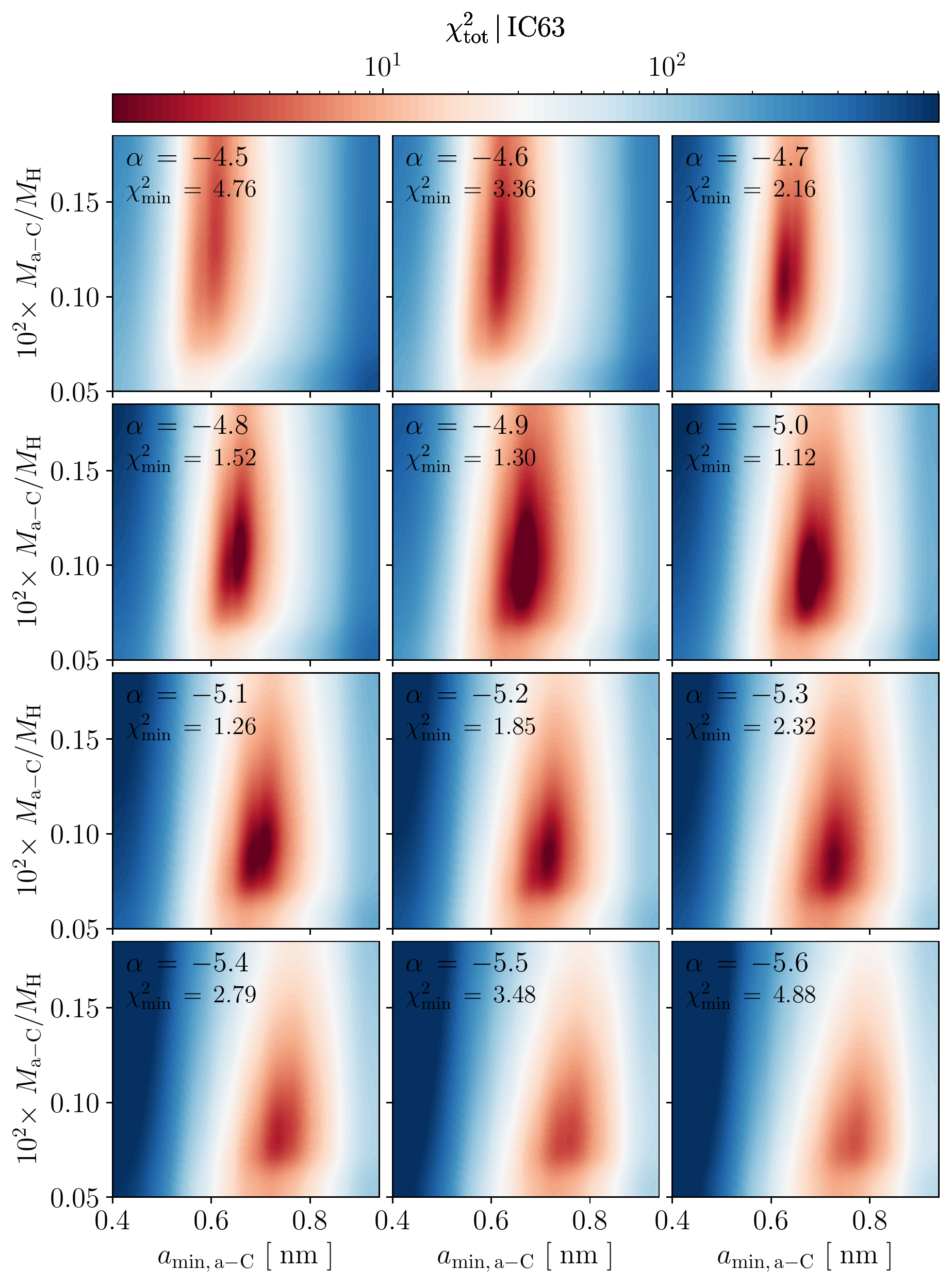}
\caption{\chiTOT~in the 2D space (\abvsg, \aminvsg). Each case corresponds to a different value of $\alpha$. The black dashed lines locate the minimum value of \chiTOT.}
    \label{fig:SOC_grid_cut_1_n0_1e5}
\end{figure*}

\section{Comparison between the dust modelled and observed emission in the Orion Bar}

\begin{figure*}[h]
\centering
        \includegraphics[width=\textwidth, trim={0 0cm 0cm 0cm},clip]{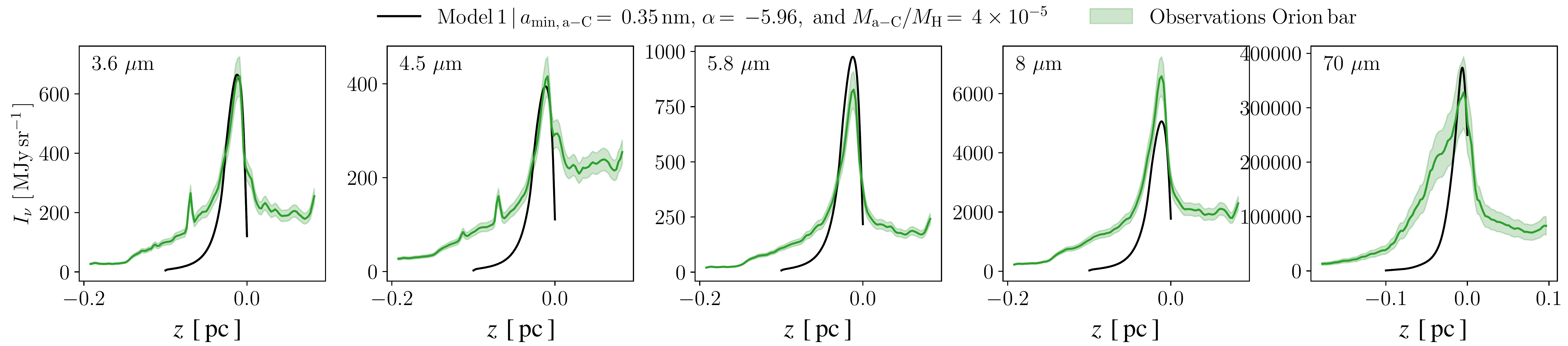}
        \includegraphics[width=\textwidth, trim={0 0cm 0cm 0cm},clip]{Orion2.pdf}
        \includegraphics[width=\textwidth, trim={0 0cm 0cm 0cm},clip]{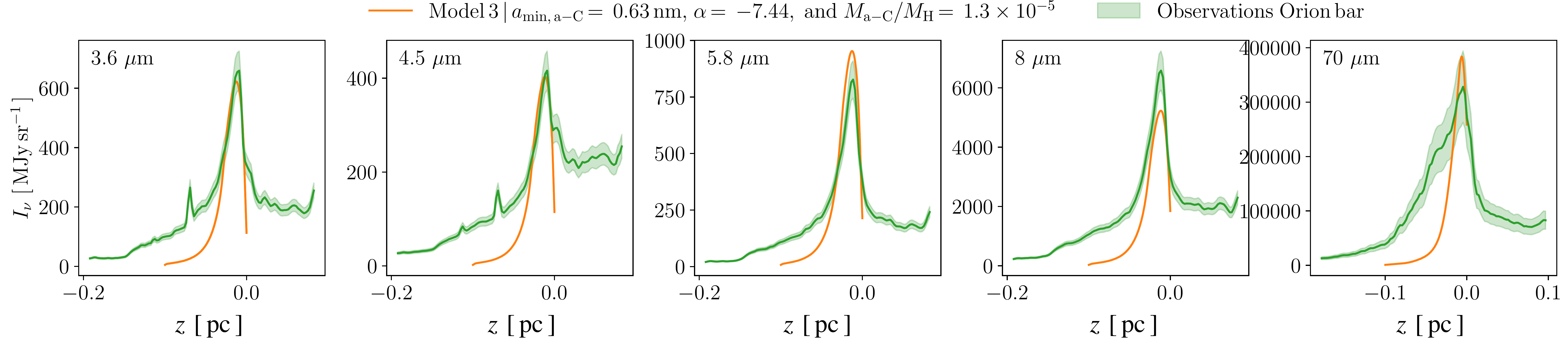}
        \includegraphics[width=\textwidth, trim={0 0cm 0cm 0cm},clip]{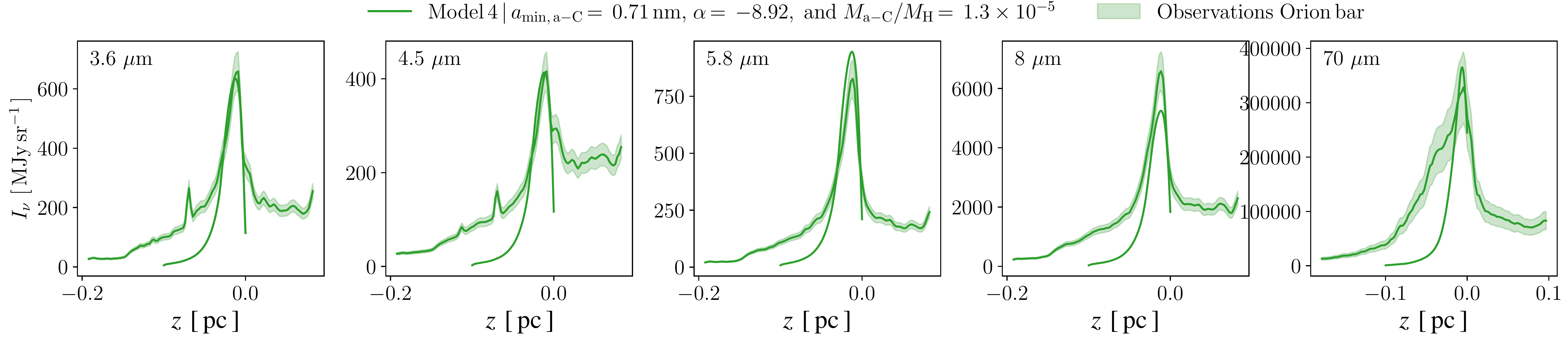}
        \includegraphics[width=\textwidth, trim={0 0cm 0cm 0cm},clip]{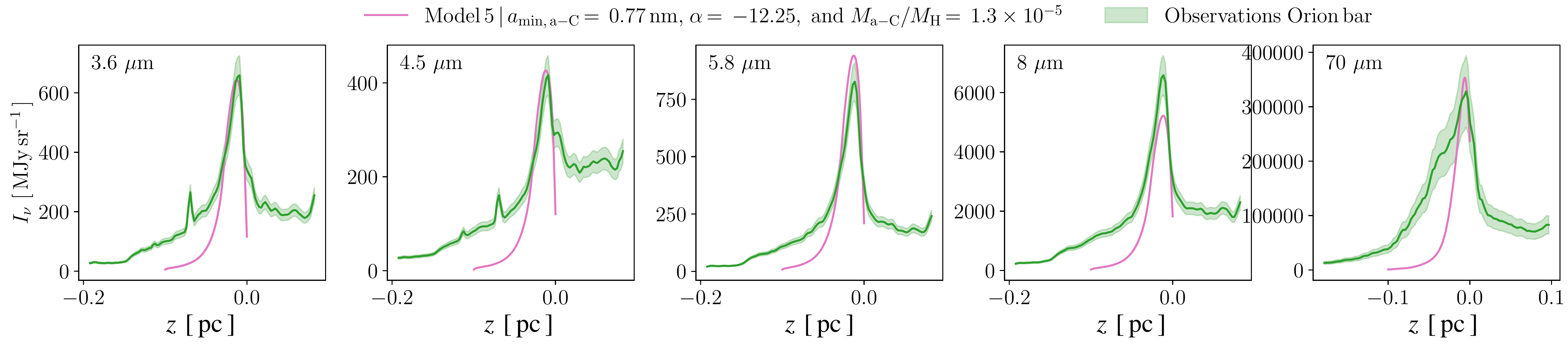}
\caption{Comparison between the observed dust emission and the modelled dust emission using the best set of dust parameters (from model 1 on the top to model 5 on bottom, showed in Fig.\,\ref{fig:orion_chi2_amin_alpha}) in five photometric bands (3.6, 4.5, 5.8, 8, and 70 \mum). The dust observed emission is shown in green lines. The cut considered across the Orion Bar is shown in Fig.\,\ref{fig:IC63_orion_obs_all}.}
    \label{fig:SOC_final_cut_orion}
\end{figure*}

\section{NIR to FIR ratio across the Orion bar}

\label{appendix:near2farIRratio}

\begin{figure}[h]
\centering
        \includegraphics[width=0.5\textwidth, trim={0 0cm 0cm 0cm},clip]{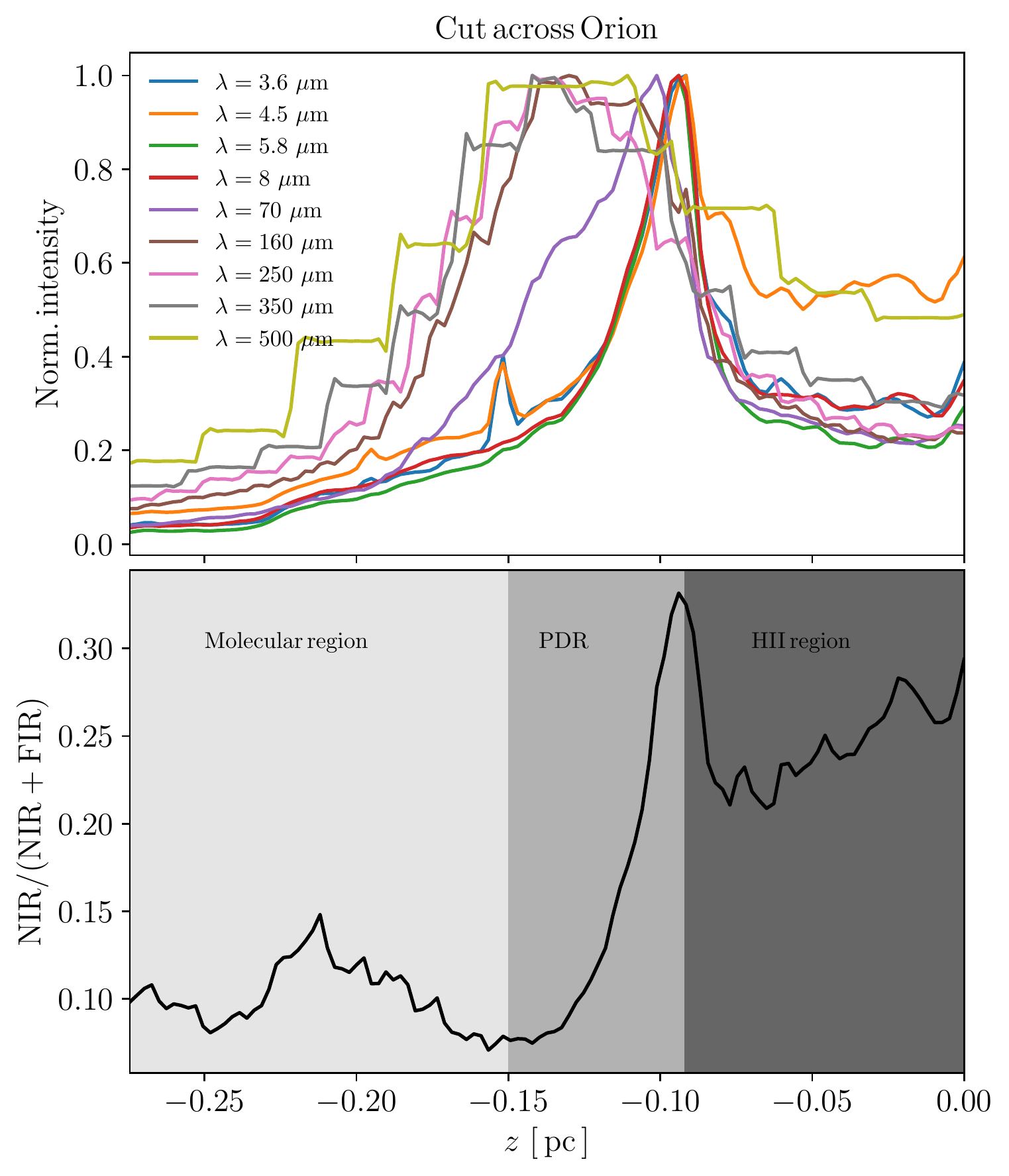}
\caption{Top: Normalised emission across a selected cut in the Orion Bar (This cut is shown in Fig.\,\ref{fig:IC63_orion_obs_all0}) in nine photometric bands. Botton: NIR to FIR\ emission ratio across the Orion Bar. The three different zones correspond to the molecular region, the PDR, and the HII regions respectively.}
    \label{fig:orion_cut}
\end{figure}

Here, we study the nano-grain to large grain abundance ratio across the Orion Bar via the NIR-to-FIR emission ratio. This approach purely observational-based allows us to qualitatively follow dust evolution without being model-dependent. We show in Fig.\,\ref{fig:orion_cut} the normalised intensities in nine photometric bands (top panel) and the NIR-to-FIR ratio (bottom panel) across a selected cut in the Orion Bar (this cut is shown in Fig.\,\ref{fig:IC63_orion_obs_all0}). 

We face a challenge here as the spatial resolution of the NIR observations with Spitzer is much better than that of the RIR observations with Herschel. In fact, while we have many points available within the PDR with Spitzer, we only have around three points from the HII regions via the PDR to the dense region. It is therefore challenging to conclude on the nano-grain to large grain abundance ratio as a function of depth inside the PDR. Despite this difficulty, we do, however, see a net trend as this ratio decreases in the PDR region when we move away from the illuminating star, which is in agreement with a decrease in the nano-grain abundance with depth inside the PDR. 

This study is limited by the fact that the spatial resolution of the NIR emission is much better than the one of the RIR emission. However, to end  on a positive note, it is thanks to JWST that there is indeed a way to follow nano-grains and large grains with the same spatial resolution: large grains emit in the FIR but scatter light in the visible/NIR. Therefore, we normally would be able to follow large grains through scattering in the visible/NIR with almost the same spatial resolution than the nano-grains through their NIR/MIR emission. This would hopefully provide us with an accurate nano-grain to large grain abundance ratio as a function of depth inside PDRs in the near future.

\section{Uncertainties on the advection timescale}
\label{app:uncertainties}

The uncertainty on the advection timescale is defined as:

\begin{equation}
    \frac{\Delta \tau_{\mathrm{ad}}}{\tau_{\mathrm{ad}}} = \sqrt{\left(\frac{\Delta L}{L}\right)^2 + \left(\frac{\Delta n_{\mathrm{H}}}{n_{\mathrm{H}}}\right)^2 + \left(\frac{\Delta G_0}{G_0}\right)^2},
\end{equation}

\noindent where the uncertainties on $L$, $n_{\mathrm{H}}$, and $G_0$ are defined in Table\,\ref{tab:resume1}.

\end{appendix}

%
%

\end{document}